\documentclass[10pt]{article}
\usepackage{upgreek}
\usepackage[mathscr]{euscript}
\usepackage{amsmath,environ}
\usepackage{amsfonts} 
\usepackage{amssymb}
\usepackage{mathrsfs}
\usepackage{graphicx,psfrag,color}
\usepackage{graphicx,amsmath,amssymb}
\usepackage{dcolumn}
\usepackage{bm}
\usepackage{graphicx,psfrag,color}
\usepackage{xcolor}
\usepackage[colorlinks=true,linkcolor=brown, citecolor=blue, urlcolor=purple]{hyperref}
\usepackage[top = 1.5in, left = 1in, bottom = 1in, right=1in]{geometry}

\usepackage{stmaryrd}
\usepackage{mathrsfs}
\usepackage{yfonts}
\usepackage{latexsym, epsf}
\usepackage{cancel}
\usepackage{amsbsy}

\usepackage{amssymb}
\usepackage{amsmath}
\usepackage{epsfig} 
\usepackage{epstopdf} 
\usepackage{graphicx}

\usepackage[firstinits=true, parentracker=true, 
doi=true, isbn=false, url=true, natbib=true,
eprint=false,sorting=nyt,]{biblatex}

\begin{document}
\large

\title{A Family of Solitary Waves in Frenkel-Kontorova Lattice\thanks{This work was completed in the period Oct 2003-Aug 2004 as part of the author's PhD at Cornell University. The author thanks Phoebus Rosakis, Anna Vainchtein, and Lev Truskinovsky for comments and suggestions. This work was partially supported by NSF grant DMS-0072514. Since the manipulations and observations presented in this paper are still anew it was decided by the author to place it in open domain.}}
\author{Basant Lal Sharma\thanks{(bs72@cornell.edu) Department of Theoretical and Applied Mechanics, Cornell University, Ithaca, NY 14853. 
Present address: {(bls@iitk.ac.in) Department of Mechanical Engineering, Indian Institute of Technology Kanpur, Kanpur, U. P. 208016, India}}}

\date{Aug 2004}

\maketitle

\begin{abstract}
A family of solitary waves is constructed in Frenkel-Kontorova model and its continuum and quasicontinuum approximations. Each solitary waves is characterised by the number of local maxima in its profile and a relation between external force and the velocity of wave. Such waves may be interpreted as a coherent motion, with constant velocity, of two dislocations of opposite sign or of kink-antikink pair.
\end{abstract}


\setcounter{section}{-1}
\section{Introduction}
\label{section1}

Plasticity and dislocations have been the focus of active research for more than a century. After a scheme was presented by \citet{disloc_prandtl} and \citet{disloc_dehlinger}, a one-dimensional model of lattice was suggested by \citet{disloc_FK} to understand plasticity in crystals. In these pioneering investigations and others, mainly dynamics of a dislocation is considered and results are obtained in the long wavelength limit \citep{seeger5, frank5, seeg_schill, seeger}. The Frenkel-Kontorova lattice model, as a Hamiltonian system, involves an onsite potential with more than one energy wells \citep{disloc_FK} or, as called in the paper, different phases. Thus, at least two types of travelling waves can be studied:
\begin{description}
\item[ Type $1$:]This is the case when particles at $\pm\infty$ are in different phases. This travelling wave may be interpreted as dislocation or kink and has been extensively studied before in the context of plasticity and phase transformation (e.g. \citet{disloc_atkinson}, \citet{earmme}, \citet{rohan1}, \citet{kresse}, \citet{caprio}), where it is commonly known as Frenkel-Kontorova dislocation. Due to the presence of external driving force, there is radiation of elastic waves.
\item[ Type $2$:]In this case, except some particles that are in different phases those at $\pm\infty$ are in the same phase. This configuration can be associated with a uniformly moving dislocation dipole or kink-antikink pair \citep{seeg_schill}.
\end{description}
In long wavelength limit of Frenkel-Kontorova model, traditionally known as sine-Gordon equation, the type $1$ configuration appears as Heteroclinic orbit, and such configuration is sometimes called sine-Gordon kink or soliton. In the same limit, the type $2$ configuration appears as Homoclinic orbit where it is also called sine-Gordon kink-antikink pair. The creation of kink-antikink pair in a dislocation in the presence of an applied stress and thermal fluctuations was treated by \citet{seeg_schill}. Recent numerical experiments by \citet{dmitriev1,dmitriev2} have demonstrated the possibility of the creation of a kink-antikink pair in the Frenkel-Kontorova model due to the interaction of two breathers \citep{kiv_mal} in the absence of a driving force. For a two dimensional Frenkel-Kontorova model at non-zero temperature, nucleation and propagation of kink-antikink pair on a dislocation has been demonstrated through computer simulations by \citet{gorno_kats_krav_tref}. However, to our knowledge a plethora of such type $2$ waves and the corresponding family of force-velocity relations \citep{trusk, rohan2} have not been discussed in the literature on Frenkel-Kontorova lattice model \citep{braun_kivshar};
a connection with Yoffe's problem is anticipated \citet{Yoffe1951,Georgiadis}.

In this paper we discuss type $2$ waves only. We assume that all particles in the lattice, except a few localised within a finite neighbourhood of second phase, are in one phase. The conventional Frenkel-Kontorova model with a periodic onsite potential can be thus replaced by a model with double well potential. \citet{sanders2} and \citet{disloc_atkinson} presented the Frenkel-Kontorova lattice model for a special choice of onsite potential that allowed representation of type $1$ waves in closed form. The choice of onsite potential in this paper is same as that of \citet{disloc_atkinson}. As the situation discussed here is similar to the juxtaposition of two dislocations with opposite sign, we may refer to the dislocation on right side as first dislocation and the other dislocation as second dislocation. These solitary waves exhibit a coherence that is reflected in wave profile as well as relation between force and velocity and persist in continuum and quasicontinuum \citep{kunin} approximations of the lattice model.

This paper is organized as follows. In the next section we formulate the one-dimensional lattice model and present the Euler-Lagrange equations. In second section we present a general expression for family of the solitary waves in the lattice model and corresponding force-velocity relation. Then we present the same for continuum and quasicontinuum approximations that capture type $2$ waves and finally we conclude with some remarks.

\section{Lattice Model}
\label{section2}

Consider a one-dimensional lattice of particles. Let the set of integers, denoted by $\mathbb{Z},$ be identified with the particles constituting the lattice. Let $\tilde{u}_{n}(t)$ denote the displacement of $n$th particle, which is located at position $n\varepsilon$ in the lattice, for each $n\in \mathbb{Z}$ and $t\in \mathbb{R}$. By $\varepsilon$ we denote the lattice parameter. Suppose the lattice is attached to a rigid foundation by bistable springs with energy density $w$ such that $w(0)=w(a)=0, w'(0)=w'(a)=0,$ and $w''(0)=w''(a)=c>0,$ for some $a>0.$ Further assume that each particle interacts with only its nearest neighbour particle through harmonic forces captured by elastic modulus $E$. This means that $n$th particle experiences the force due to potential energy $\varepsilon \tilde w(\tilde{u}_n),$ in addition to the force due to harmonic potential involving the discrepancy with the displacements of its nearest neighbours, $\tilde{u}_{n-1}$ and $\tilde{u}_{n+1}$. Finally, we introduce an external force, per unit length and independent of $n$, on each particle and call it $\tilde\sigma.$ We use the Lagrangian
$$\mathfrak{L}(\{\tilde{u}_i\}_{i\in\mathbb{Z}})=\sum_{n\in\mathbb{Z}}\frac{1}{2}\rho\varepsilon \dot{\tilde{u}}_n^2-\sum_{n\in\mathbb{Z}}\{\frac{1}{2}E\varepsilon((\tilde{u}_{n+1}-\tilde{u}_{n})/{\varepsilon})^2+\varepsilon \tilde w(\tilde{u}_{n})-\varepsilon \tilde\sigma\tilde{u}_n\}$$
where $\rho $ is the mass density per unit length along the lattice, $\dot{\tilde u}=d{\tilde u}_n(t)/dt$ and we suppress the dependence on $t$ in these expressions. The Euler-Lagrange equation describing the motion for $n$th particle is
$$
\rho \varepsilon \ddot{\tilde{u}}_{n}-\frac{E}{\varepsilon }
(\tilde{u}_{n+1}-2\tilde{u}_{n}+\tilde{u}_{n-1})+\varepsilon \lbrack \tilde w^{\prime }(\tilde{u}_{n})-\tilde\sigma ]=0, \forall n\in\mathbb{Z}.
$$

Let $t=\tilde t/(\varepsilon /\sqrt{E/\rho }), u_{n}=(2\tilde u_{n}/a-1), \sigma=\tilde\sigma /(ac)$ and $w(u_n)=2/(ca^2)\tilde w(\tilde u_n).$ Let $x=\tilde x/\varepsilon$ describe the one-dimensional lattice in physical space. The Euler-Lagrange equations, in the dimensionless formulation of the lattice model, can be rewritten as
\begin{equation}
\ddot{u}_{n}-(u_{n+1}-2u_{n}+u_{n-1})+\chi^{2}[w^{\prime }(u_{n})-\sigma ]=0, \forall n\in\mathbb{Z},
\label{discrete model}
\end{equation}
where $\chi=\varepsilon \sqrt{2c/{E}}$ is a structural parameter.
\begin{figure}[h]
\begin{center}
\includegraphics[height=4in]{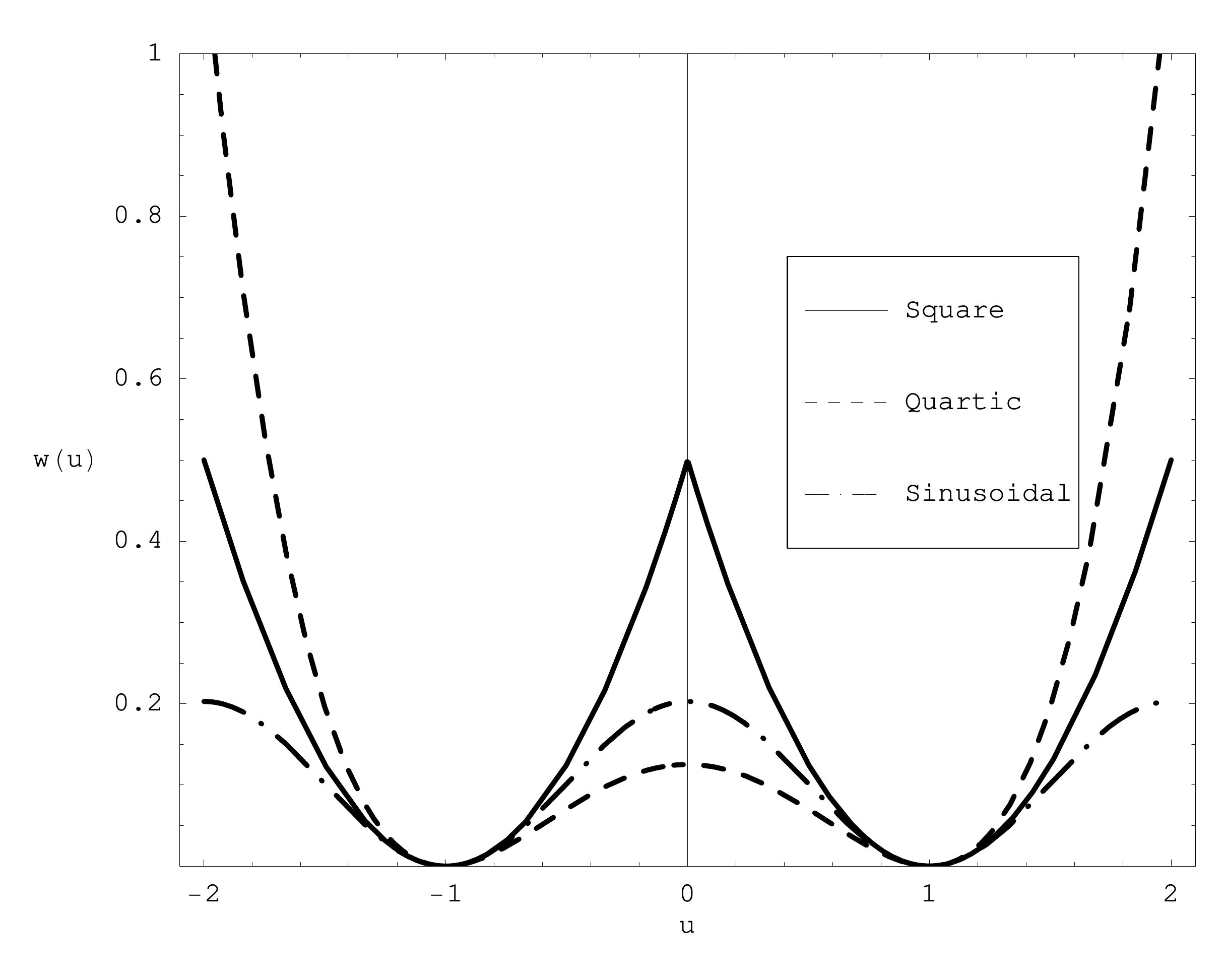}
\end{center}
\caption{Bi-stable Potential}
\label{potential}
\end{figure}
The function $w$ has global minima at $\pm 1$ and also a local maxima at $0$. By the phrase `$n$th particle in the lattice is in the first (second) phase', it is meant that $u_n<0 (>0).$ We consider a special potential function for $w$, so that we can present solutions in closed form,
\begin{equation}
w(u)=\frac{1}{2}(u+1)^{2}H(-u)+\frac{1}{2}(u-1)^{2}H(u),
\label{square}
\end{equation}
where $H(x)=1$ for $x\ge0$ and $0$ otherwise is the Heaviside step function. This square well potential function is also shown in Fig. \ref{potential} alongwith a quartic well potential function $w(u)=\frac{1}{8}(u-1)^{2}(u+1)^{2}$ and sinusoidal potential function $w(u)=\frac{1}{\pi^2}(1+\cos\pi x)$ with the same curvature at $\pm1.$ In the context of dislocation, an approximation of a nonlinear function by piecewise linear function has been used by \citet{maradudin1}, \citet{disloc_sanders}, \citet{disloc_celli} and \citet{disloc_ishioka} for a screw dislocation, and by \citet{disloc_krato_inden}, \citet{sanders2}, and \citet{disloc_atkinson} for Frenkel-Kontorova model.

In general, one would like to investigate the behaviour of $\{u_n(t)\}_{n\in\mathbb{Z}}$ with given initial and boundary conditions as well as given force $\sigma$ and parameter $\chi.$ However, in this paper we discuss only solitary waves.

\section{Solitary Wave in Lattice Model}
\label{section3}

Let $z=n-vt,$ $u(z)=u_{n}(t)$ in (\ref{discrete model}) where $v$ is the velocity of solitary wave. The piecewise linear differential-difference equation that
determines $u(z)$ is
\begin{equation}
v^{2}d^{2}u/dz^{2}-(u(z+1)-2u(z)+u(z-1))+\chi^{2}[u(z)-{\sigma+1}-2H(u)]=0.
\label{disc_steady}
\end{equation}
Assume that
\begin{equation}
u(z)<0, \left| z\right| >z_{0}; u(z)>0, \left| z\right| <z_{0}.
\label{disc_ans1}
\end{equation}
Also we assume that $u(z)\rightarrow$ const. as $\left| z\right| \rightarrow \infty.$ The assumption \eqref{disc_ans1}, similar to that used by \citet{disloc_atkinson}, \citet{earmme} and \citet{ kresse} owing to the special choice of onsite potential function, leads to the following {\em linear} differential-difference equation
\begin{equation}
v^{2}d^{2}u/dz^{2}-(u(z+1)-2u(z)+u(z-1))+\chi^{2}[u(z)-{\sigma+1}-2H(z_0-\left| z\right|)]=0.
\label{disc_steady2}
\end{equation}
By using Fourier Transforms, we can construct the solitary wave\footnote{For details please see appendix \ref{app_dynamics}.} from equation (\ref{disc_steady2})
\begin{equation}
u(z)={\sigma-1}+2\chi^{2}\left\{ 
\begin{array}{cc}
2i\sum_{\xi \in S_{d}^{+}}e^{-i\xi z}\sin (\xi z_{0})/(L^{\prime }(\xi)\xi)& z\leq -z_{0}, \\ 
\left[ 
\begin{array}{c}
1/\chi^{2}+\sum_{\xi \in S_{d}^{+}}e^{-i\xi z}e^{i\xi z_{0}}/(L^{\prime }(\xi )\xi) \\ 
+\sum_{\xi \in S_{u}}{e^{-i\xi z}e^{i\xi z_{0}}}/(L^{\prime }(\xi )\xi)\\
+\sum_{\xi \in S_{d}^{-}}e^{-i\xi z}e^{-i\xi z_{0}}/(L^{\prime }(\xi )\xi)
\end{array}
\right. & \left| z\right| <z_{0}, \\ 
-2i\sum_{\xi \in S_{d}^{-}}{e^{-i\xi z}\sin (\xi z_{0})}/(L^{\prime }(\xi )\xi) & z\geq z_{0},
\end{array}
\right. \label{disc_sol}
\end{equation}
with $L(\xi )=\chi^{2}+4\sin^{2}{\xi}/{2}-v^{2}\xi^{2}, S_{d}^{\pm }=\{\xi \mid L(\xi )=0;\text{Im }\xi \gtrless 0\}, S_{u}=\{\xi \mid L(\xi )=0;\text{Im }\xi=0, \xi\neq0\}.$ By assumption \eqref{disc_ans1}, using $u(z_{0})=0,$ the constant force $\sigma$ is given by
\begin{equation}
\sigma=1+4\chi^{2}i\sum_{\xi \in S_{d}^{-}}e^{-i\xi z_{0}}\sin(\xi z_{0})/(L^{\prime }(\xi )\xi)
\label{disc_u0}
\end{equation}
and by using $u(-z_0)=0$, $z_{0}$ is determined by
\begin{equation}
\sum_{k\in S_{u}\cap \mathbb{R}^{+}}\sin^{2}kz_{0}/(L^{\prime }(k)k)=0.
\label{def_z0}
\end{equation}
Let us define $u(z_0^+)=\lim_{z\searrow z_0}u(z), u(z_0^-)=\lim_{z\nearrow z_0}u(z).$ We can show\footnote{For details please see Claim 3 in appendix \ref{app_dynamics}.} that $u^{\prime }(z_{0}^{-})=u^{\prime }(z_{0}^{+}), u^{\prime }(-z_{0}^{+})=4\chi^{2}\sum_{\xi \in S_{u}\cap \mathbb{R}^{+}}\sin (2kz_{0})/L^{\prime }(k)-u^{\prime }(z_{0}^{+}),u^{\prime }(-z_{0}^{-})=-u^{\prime }(z_{0}^{+}),$ therefore
\begin{equation}
u^{\prime }(-z_{0}^{+})=-u^{\prime }(-z_{0}^{-})\Leftrightarrow\sum_{\xi \in S_{u}\cap \mathbb{R}^{+}}\sin (2kz_{0})/L^{\prime}(k)=0.
\label{condsmooth}
\end{equation}
The function (\ref{disc_sol}) is a solution of \eqref{disc_steady} only if the particles in the interval $(-z_{0},z_{0})$ lie in second phase according to the assumption \eqref{disc_ans1}. We present below an interval of $v$ where the solution (\ref{disc_sol}) may hold.

\begin{figure}[h]
\begin{center}
\includegraphics[height=2.4in]{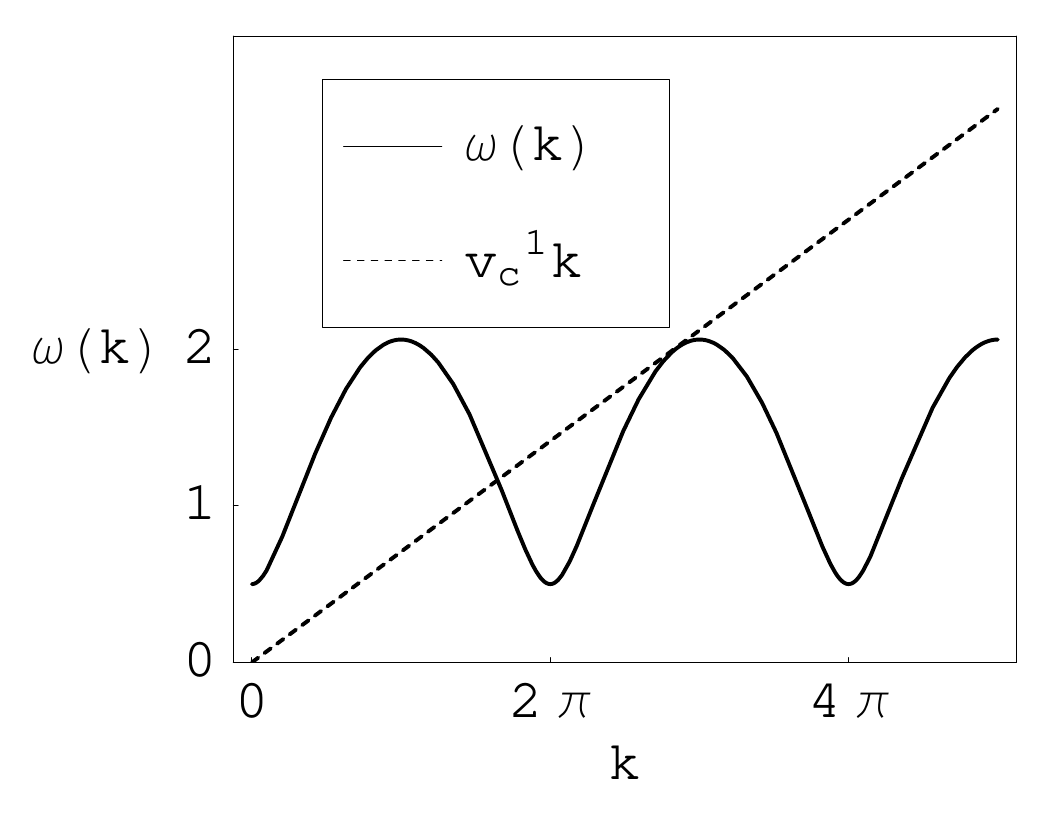}
\end{center}
\caption{$v^1_c$ for $\chi=0.5$.}
\label{vc1}
\end{figure}

When $v>v^1_c,$ where $v^1_c$ is the maximum phase velocity that equals the group velocity for the same non-zero wave number \citep{disloc_atkinson, rogula}, as shown in Fig. \ref{vc1}, we get $S_{u}=\{k_{0},-k_{0}\}.$ From (\ref{def_z0}) we obtain $z_{0}=n\pi/k_{0}, n\in \mathbb{Z}, n>0$ and the condition (\ref{condsmooth}) is automatically satisfied. Thus a {\em countable} number of solitary waves may exist for given velocity $v>v^1_c.$ For $v<v^1_c,$ the expression (\ref{disc_sol}) may not be a solution of \eqref{disc_steady} as we could not find $v$ and $z_0,$ numerically, satisfying condition (\ref{condsmooth}) and assumption \eqref{disc_ans1}, similar to the case of one dislocation \citep{disloc_atkinson, disloc_celli}. We also discuss this issue at the end of the paper.

\begin{figure}[h]
\begin{center}
\includegraphics[width=\textwidth]{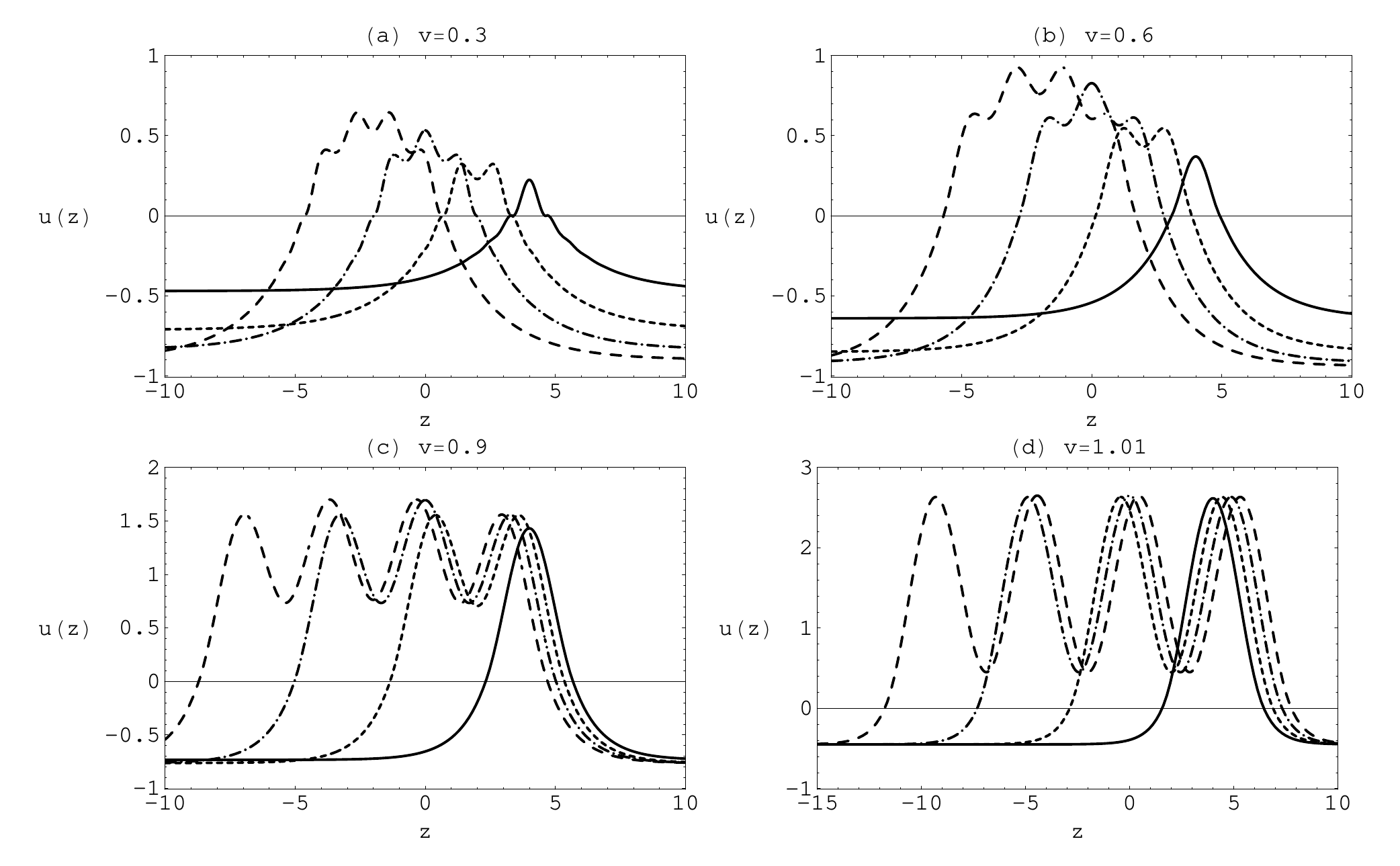}
\end{center}
\caption{Solitary waves in Frenkel-Kontorova Lattice for $\chi=0.5$. The wave profiles have been shifted a little in order distinguish them visually. The solid curves stands for $n=1,$ small dashed curve for $n=2,$ dash-dot-dash curve for $n=3,$ and long dashed curve for $n=4.$}
\label{solwaveD}
\end{figure}

For solitary waves presented here the velocity of propagation should be read as speed of propagation since the wave \eqref{disc_sol} with prescribed velocity do not depend on the sign of $v.$ As shown in Fig. \ref{solwaveD}, symbol $n$ corresponds to the number of local maxima in wave profile (\ref{disc_sol}), or {\em bumps} in the second phase. The solitary wave profile is shown in Fig. \ref{solwaveD} for one and more bumps in the region with second phase and these waves have been grouped together with respect to same velocities of propagation. At small velocities there is a notable difference in the profiles for different values of $n$ but at large velocities the wave profiles resemble each other very closely. Also at large velocities, starting around $v=1$, the displacements become so large that dislocation dissociation may be favored similar to the case of one dislocation \citep{earmme3}.

Note that the size of second phase zone, twice of $z_0$ \eqref{def_z0}, depends only on the non-zero and real wave numbers, $k\in S_u,$ satisfying the condition $L(k)=0,$ or equivalently $\omega(k)=v k, k\in\mathbb{R},$ where $\omega(k)=\chi^{2}+4\sin^{2}{k}/{2}$ is the dispersion relation for the lattice. This is reminiscent of the possiblity of macroscopic dissipation in the form of microscopic waves in the analysis of Frenkel Kontorova dislocation \citep{disloc_atkinson, kresse, caprio} and also in the analysis of a screw dislocation \citep{disloc_celli, sharma1, sharma2}, where a dislocation moves due to the action of a configurational force and work done in this way at macroscale is converted into energy radiated in microscopic waves. It can be observed that, in case of one dislocation, use of inverse radiation condition leads to motion of dislocation against the configurational force. In other words, the energy contained in waves is converted into potential energy of configurational type and this can be interpreted as non-dissipative motion.

In fact, in each solitary wave, energy (in waves) is radiated by the first dislocation and it is macroscopically dissipative and the second dislocation with opposite sign is activated by these radiated waves to move against the action of configurational force and is non-dissipative. Taking sum of both there is no dissipation and also there is no work done by configuration forces when the dislocation dipole moves. Indeed, one interpretation of these solitary waves is that they represent two dislocations of opposite sign moving together in such {\em coherence} that one dislocation emits lattice waves in a manner that supplies the right condition for the propagation of the other dislocation\footnote{This comment is due to Lev Truskinovsky.}. This is a much stronger interaction between waves and dislocation than that studied by \citet{leib3}, \citet{nabarro2} and \citet{esh4} where a wave is considered as a small perturbation leading to oscillatory motion of a dislocation.

If the two dislocations are separated by a large distance, i.e. $n\rightarrow\infty$, then the force needed to move second dislocation is independent of the other and in this limiting case $\sigma v$-relation \eqref{disc_u0} is same as the kinetic relation \citep{trusk, rohan2} for one dislocation obtained by \citet{disloc_atkinson}. Therefore, we can interpret the relation \eqref{disc_u0} between external force and velocity as the kinetic relation for first dislocation {\em modified} by the presence of the second dislocation of opposite sign moving with the same velocity. For simplicity, we call the relation between force, $\sigma,$ and velocity, $v,$ of a solitary wave as {\em $\sigma v$-relation}. For the case $v>v^1_c,$ $S_u=\{k_0, -k_0\}$ and $\sigma v$-relation \eqref{disc_u0} reduces to
$$\sigma=2\chi^{2}\sum_{\xi\in S_{d}^{-}}(1-e^{-2i\xi n\pi/k_{0}})/(L^{\prime }(\xi )\xi).$$
\begin{figure}[h]
\begin{center}
\includegraphics[height=3.5in]{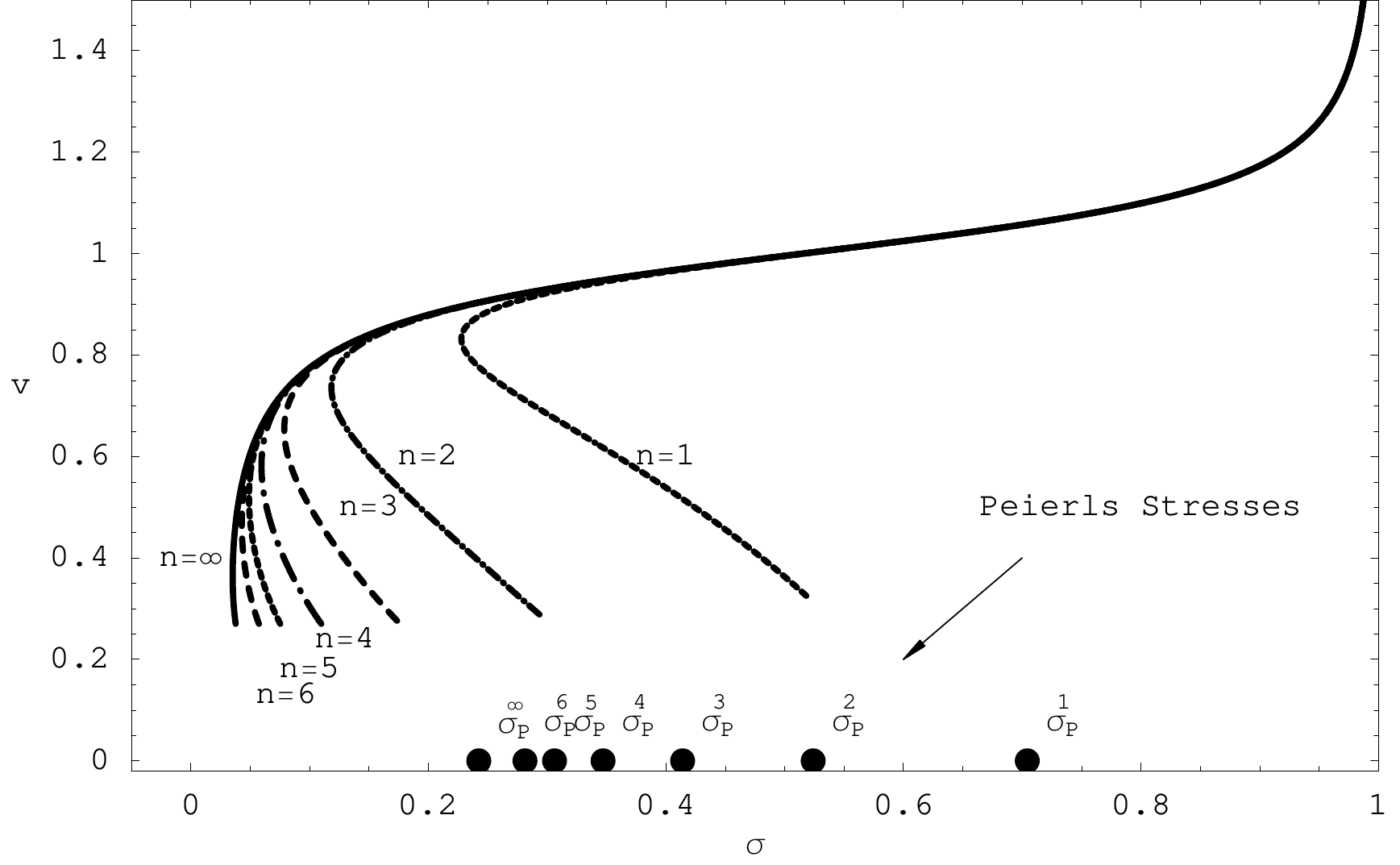}
\end{center}
\caption{$\sigma v$-relations for solitary waves in Frenkel-Kontorova lattice for $\chi=0.5.$ Peierls stresses are obtained in \citep{sharma5}}
\label{krd}
\end{figure}
Thus there is a unique stress level $\sigma$ associated with the propagation of a solitary wave with a given number of bumps $n$ and a given velocity $v$. We present $\sigma v$-relations in Fig. \ref{krd} alongwith the Peierls stress reported by \citet{sharma5}. As mentioned above, $\sigma v$-relation of type $2$ waves approaches that for type $1$ waves as $n$ increases to infinity. Note also that $\sigma$ is almost independent of $n$ for $v\ge1.$ This is resonant with resemblance in the wave profiles at large velocities in Fig. \ref{solwaveD}. This is because `core' of both dislocations shrinks at large velocity and therefore, the effective distance between them also becomes large and the resulting $\sigma v$-relations coincide with the kinetic relation for one dislocation.

In the following section, we construct the solitary wave in continuum and quasicontinuum approximations of Frenkel-Kontorova lattice model \citep{kresse}.

\section{Solitary Wave in Continuum and Quasicontinuum Approximations of Lattice Model}
\label{section5}
The phonon dispersion relation for the lattice model is given by
\begin{equation}
\omega(k)^{2}=4\sin^{2}\frac{k}{2}+\chi^{2}, \forall k\in\mathbb{C}.
\label{discrete dispersion}
\end{equation}
As the first approximation of lattice model, we get a dispersion relation for the classical continuum model
\begin{equation}
\omega^{2}=k^{2}+\chi^{2}+o(k^2),
\label{classical disp}
\end{equation}
as $k\rightarrow0.$ This corresponds to equation of motion,
\begin{equation}
u_{tt}=u_{xx}-\chi^{2}[w^{\prime }(u)-\sigma ],
\label{classical model}
\end{equation}
and reduces to the ubiquitous sine-Gordon equation \citep{seeger5, kiv_mal} or Enneper equation \citep{seeger} in a special case of $\sigma=0$ and sinusoidal potential function $w(u)=\frac{1}{\pi^2}(1+\cos\pi x).$ Using $u(x,t)=u(x-vt)=u(z)$ and the boundary conditions, $\lim_{z\rightarrow\infty }u(z)=\lim_{z\rightarrow-\infty }u(z),$ it can be found that there are no type $2$ solitary waves for $\sigma=0,$ but for $\sigma>0,$
$$\pm \int_{u(0)}^{u(z)}du/\{w(u)-\{w({\sigma-1})+w^{\prime}({\sigma-1})(u-{\sigma-1})\}\}^{1/2}=\chi\sqrt{2}z/{\sqrt{1-v^{2}}}.$$
\begin{figure}[h]
\begin{center}
\includegraphics[height=3.5in]{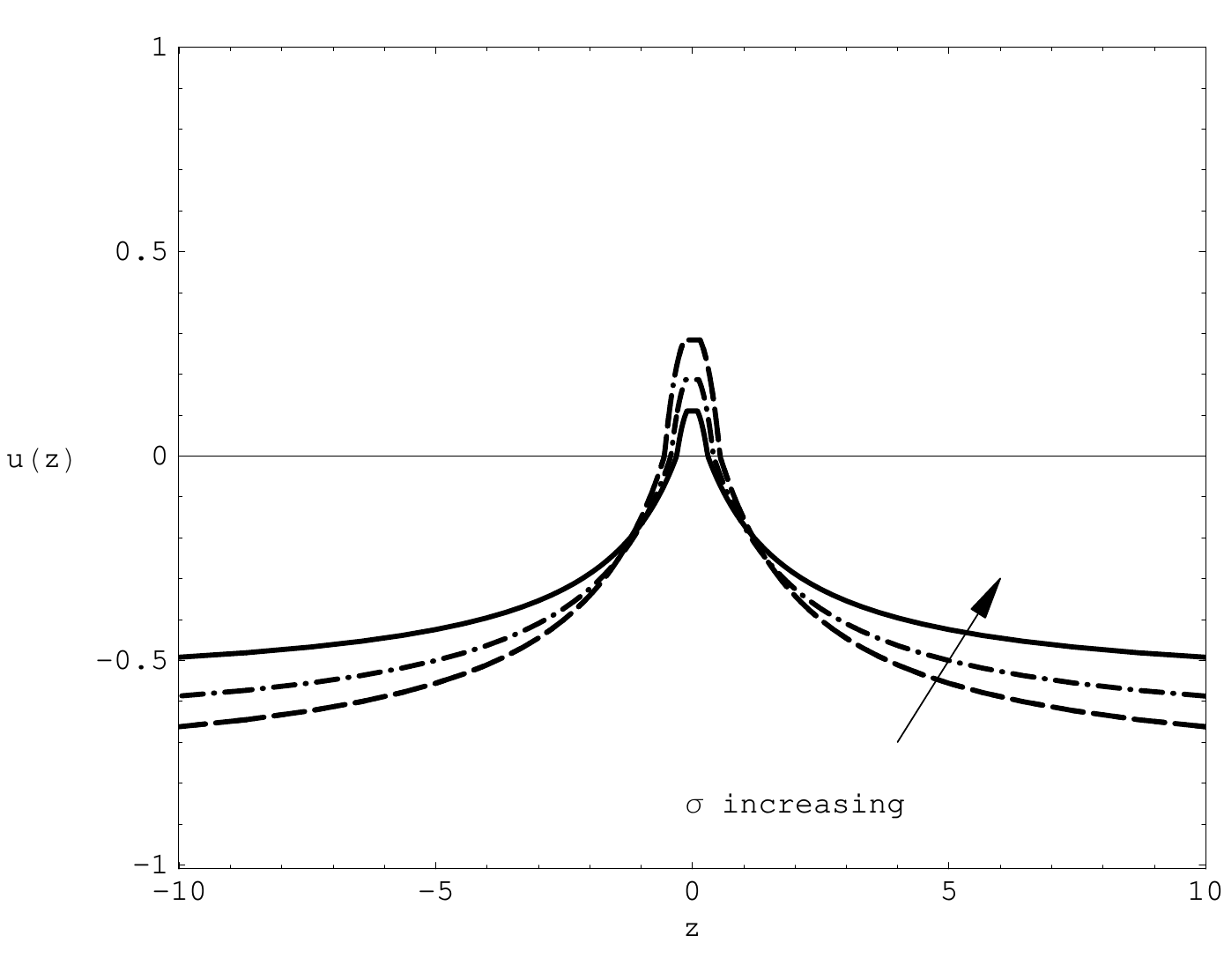}
\end{center}
\caption{Solitary waves in model $0$.}
\label{solwaveC}
\end{figure}
The solitary wave for $w$ given by \eqref{square} is shown in Fig. \ref{solwaveC} where $z-$axis has been scaled by a factor of ${\chi\sqrt{2}}/\sqrt{1-v^{2}}.$ The $\sigma v$-relation is trivial in this case, namely, any $\sigma\in(0, 1)$ is associated with any $|v|<1.$

A higher gradient continuum with negative capillarity, and not conventional positive capillarity \citep{vanderwaals, cahn_hilliard, carr_gurt_slem}, by an approximation of the lattice model is possible if we include the next term in the Taylor series expansion of the relation (\ref{discrete dispersion}). So we get a dispersion relation
\begin{equation}
\omega^{2}=k^{2}(1-\lambda k^{2})+\chi^{2}+o(k^4),
\label{seconddisp}
\end{equation}
as $k\rightarrow0,$ with\footnote{For $\lambda <0$, there are no type $2$ waves. Besides this value, $\lambda$ can be also considered as a parameter \citep{sharma1}} $\lambda=1/12.$ This is a second gradient elastic model, also called Boussinesq approximation sometimes \citep{kresse, vedantam}, with the equation of motion
\begin{equation}
u_{tt}=u_{xx}+\lambda u_{xxxx}-\chi^{2}[w^{\prime }(u)-\sigma ].
\label{secondeqn}
\end{equation}
We call this model as Model $1$. Using $u(x,t)=u(x-vt)=u(z),$ the equation for travelling waves is
\begin{equation}
\lambda u_{zzzz}+(1-v^{2})u_{zz}-\chi^{2}[u(z)-{\sigma+1}-2H(u)]=0.
\label{secondsteady}
\end{equation}
\begin{figure}[h]
\begin{center}
\includegraphics[height=3in]{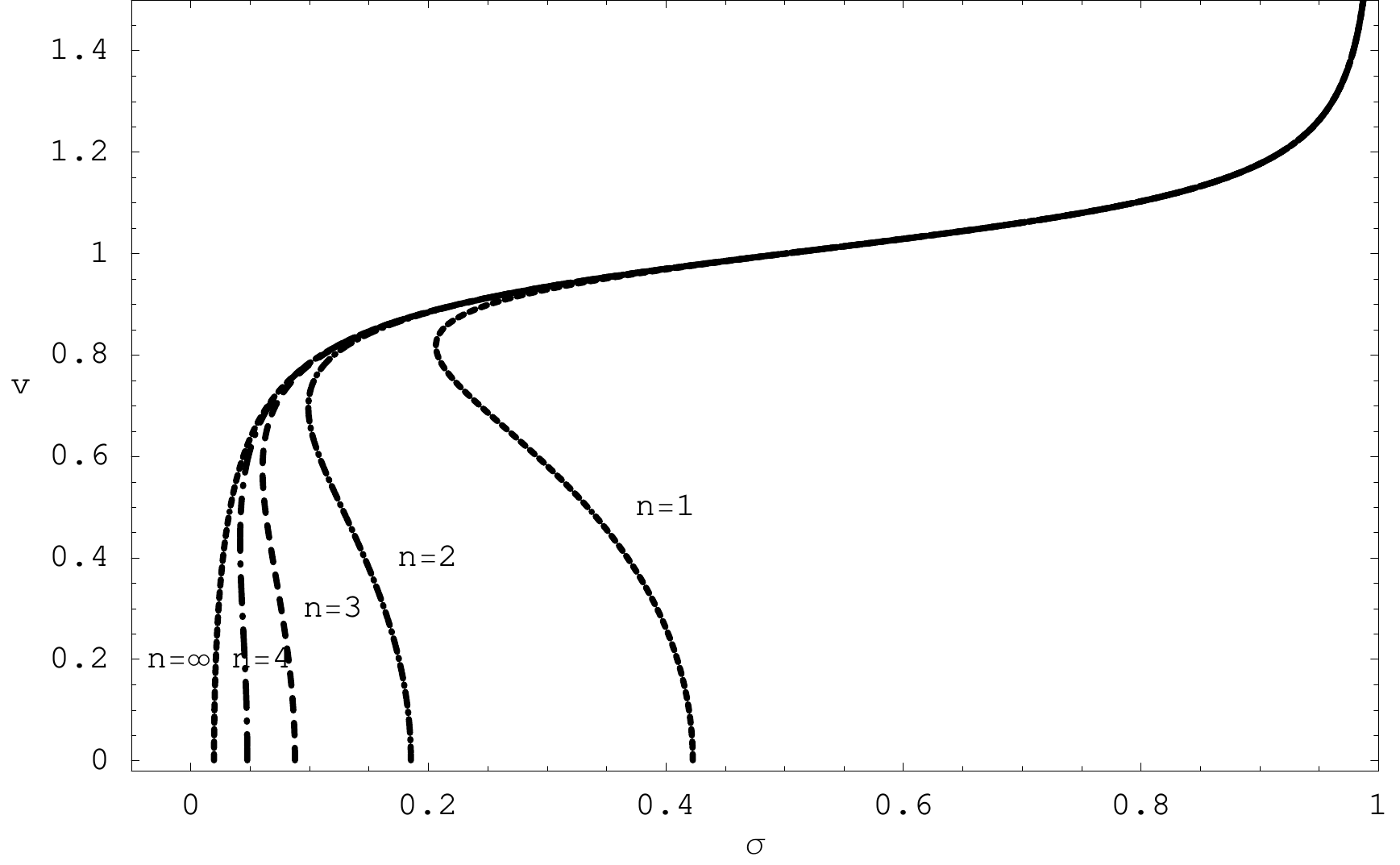}
\end{center}
\caption{$\sigma v$-relations for solitary waves in model $1$ for $\chi=0.5.$}
\label{krc2}
\end{figure} 
Assume $u>0$ for $z\in(-z_{0},$ $z_{0})$ and $u<0$ for $z\in\mathbb{R}\smallsetminus(-z_{0},$ $z_{0}).$ With $\lim_{z\rightarrow\infty }u(z)=\lim_{z\rightarrow-\infty }u(z)=$ const. and the condition that all derivatives of $u$ approach $0$ as $z\rightarrow\pm\infty,$ and the jump conditions, easily derivable from the weak form of \eqref{secondsteady}, $\llbracket u\rrbracket=\llbracket u_z\rrbracket=\llbracket u_{zz}\rrbracket=\llbracket u_{zzz}\rrbracket=0$ at $z=\pm z_0,$ the solution\footnote{The method to solve this equation is similar to the one shown in appendix \ref{app_continuum}.} can be expressed as
\begin{equation}
u(z)=\sigma-1+2\left\{ 
\begin{array}{cc}
{\sinh (n\pi r)}e^{-r_{1}\left| z\right| }/({r^{2}+1}), & z<-n\pi/r_{3}, z>{n\pi }/{r_{3}}, \\ 
\left[ 
\begin{array}{c}
1-(e^{-n\pi r}\cosh r_1z\\
+r^{2}(-1)^n\cos r_3z)/({r^{2}+1}),
\end{array}
\right. & z\in[-n\pi/ r_3, n\pi/ r_3],
\end{array}
\right. \label{sol_Grad}
\end{equation}
where $n$ is a positive integer and $r_1=\sqrt{\frac{-(1-v^{2})+R_1}{2\lambda}}, r_3=\sqrt{\frac{(1-v^{2})+R_1}{2\lambda}},$ with $R_1=\sqrt{(1-v^{2})^{2}+4\chi^{2}\lambda}$ and $r={ r_{1}}/{ r_{3}}={4\chi^{2}\lambda}/\{(1-v^{2})+R_1\},$
and $$\sigma=\frac{r^{2}+e^{-2n\pi r}}{ r^{2}+1}.$$
Using $\lambda=1/12,$ the external driving force $\sigma $ has been plotted against velocity $v$ in Fig. \ref{krc2}.

A Pad\'{e} approximation of the relation (\ref{discrete dispersion}) gives a dispersion relation representing a class of continuum models with velocity gradients \citep{mindlin1, rosenau, kresse, sharma2}
\begin{equation}
\omega^{2}=\frac{k^{2}}{1+\kappa k^{2}}+\chi^{2}, 
\label{dispersion pade22}
\end{equation}
with\footnote{$\kappa $ can also be considered as a parameter \citep{sharma2}.} $\kappa =1/{12}.$ The equation of motion for this continuum model with first order `micro-kinetic' energy \citep{theil, kresse} is
\begin{equation}
u_{tt}-\kappa u_{ttxx}=u_{xx}-\chi^{2}[w^{\prime }(u)-\sigma ]+\kappa \chi^{2}w^{\prime }(u)_{xx}.
\label{eqnpade22}
\end{equation}
We call this model as Model $2$. Using $u(x,t)=u(x-vt)=u(z),$ the equation for travelling waves is
\begin{equation}
\kappa v^{2}u_{zzzz}+(1-v^{2}+\kappa \chi^2)u_{zz}-\chi^{2}[u(z)-{\sigma+1}-2H(u)]=0.
\label{steadypade22}
\end{equation}
\begin{figure}[h]
\begin{center}
\includegraphics[height=3in]{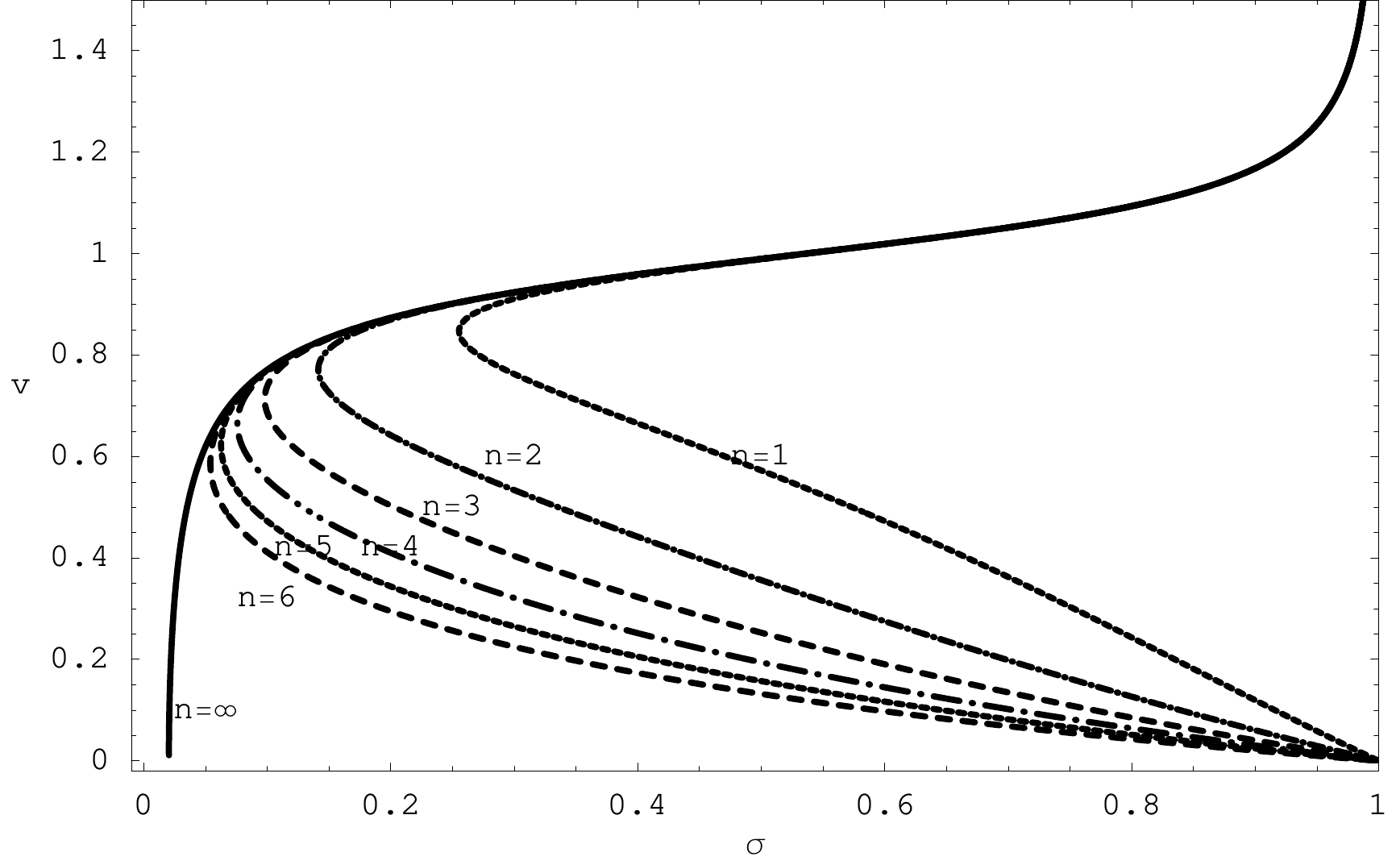}
\end{center}
\caption{$\sigma v$-relations for solitary waves in model $2$ for $\chi=0.5.$}
\label{krc}
\end{figure}
With $\lim_{z\rightarrow\infty }u(z)=\lim_{z\rightarrow-\infty }u(z)$ and the condition that all derivatives of $u$ approach $0$ as $z\rightarrow\pm\infty=$ const. and the jump conditions $\llbracket u\rrbracket=\llbracket u_z\rrbracket=\llbracket u_{zzz}\rrbracket=0, v^2\llbracket u_{zz}\rrbracket=2\chi^2$ at $z=\pm z_0,$ the solution can be written as\footnote{Please see appendix \ref{app_continuum}.}
\begin{equation}
u(z)={\sigma-1}+\left\{ 
\begin{array}{cc}
A_1e^{-r_{1}\left| z\right| }, & |z|>z_0, \\ 
2+2B_{1}\cosh{r_{1}z}+2B_{3}\cosh{r_{3}z}, & z\in \lbrack -z_{0},z_{0}]
\end{array}
\right.
\label{pade22soln}
\end{equation}
where $A_1=2\sinh{r_1z_0}\frac{\chi^2/v^2+r_3^2}{r_1^2-r_3^{2}},$ $B_1=e^{-r_1 z_0}\frac{\chi^2/v^2+r_3^2}{r_1^2-r_3^{2}},$ $B_3=-\frac{\chi^2/v^2+r_1^2}{(r_1^2-r_3^2)\cosh r_3z_0}, z_{0}=i\frac{n\pi }{r_{3}},$ with $n$ being a positive integer and $r_{1}=\sqrt{\frac{-(1-v^{2}+\kappa \chi^2)+R_2}{2\kappa v^{2}}},$ $r_{3}=i\sqrt{\frac{(1-v^{2}+\kappa \chi^2)+R_2}{2\kappa v^{2}}},$ \\
$R_2=\sqrt{(1-v^{2}+\kappa \chi^2)^{2}+4\chi^{2}\kappa v^{2}}.$ The Kinetic relation is $$\sigma=\frac{\chi^2/v^2+r_1^2+(\chi^2/v^2+r_3^2)e^{-2i n\pi r_1/r_3}}{r_1^2-r_3^{2}}.$$ Using $\kappa ={1}/{12},$ the external driving force $\sigma $ necessary to sustain the motion of the solitary waves has been plotted against velocity $v$ in Fig. \ref{krc}. For model $2$, as $v\rightarrow 0, \sigma \rightarrow 1, \forall n<\infty$ in contrast with the case of model $1$.

\section{Some Remarks}
\label{section6}

\begin{figure}[h]
\begin{center}
\includegraphics[width=\textwidth]{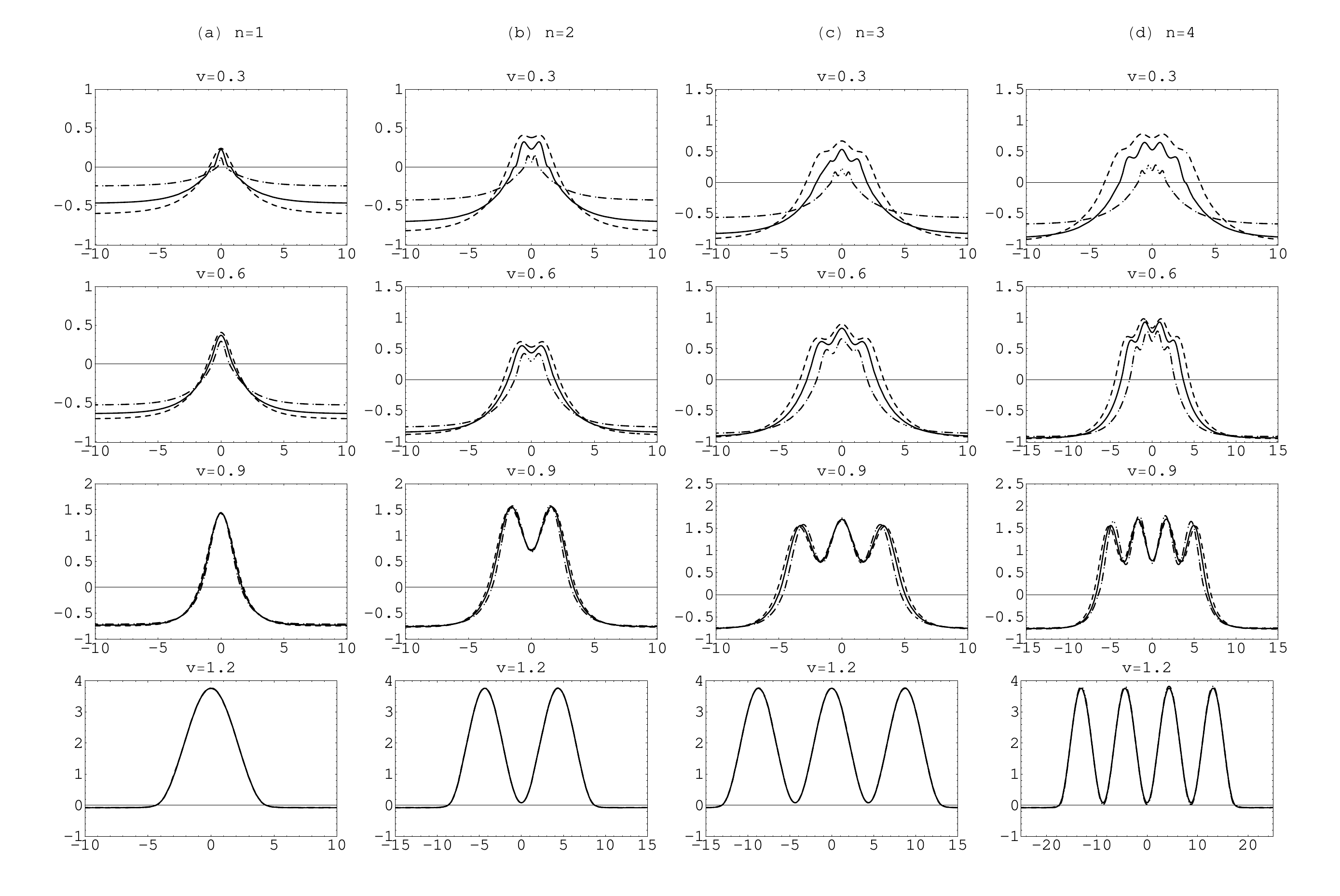}
\end{center}
\caption{Solitary waves for $n=1,2,3,4$.}
\label{solwavecompare}
\end{figure}
\begin{figure}[th]
\begin{center}
\includegraphics[height=3.5in]{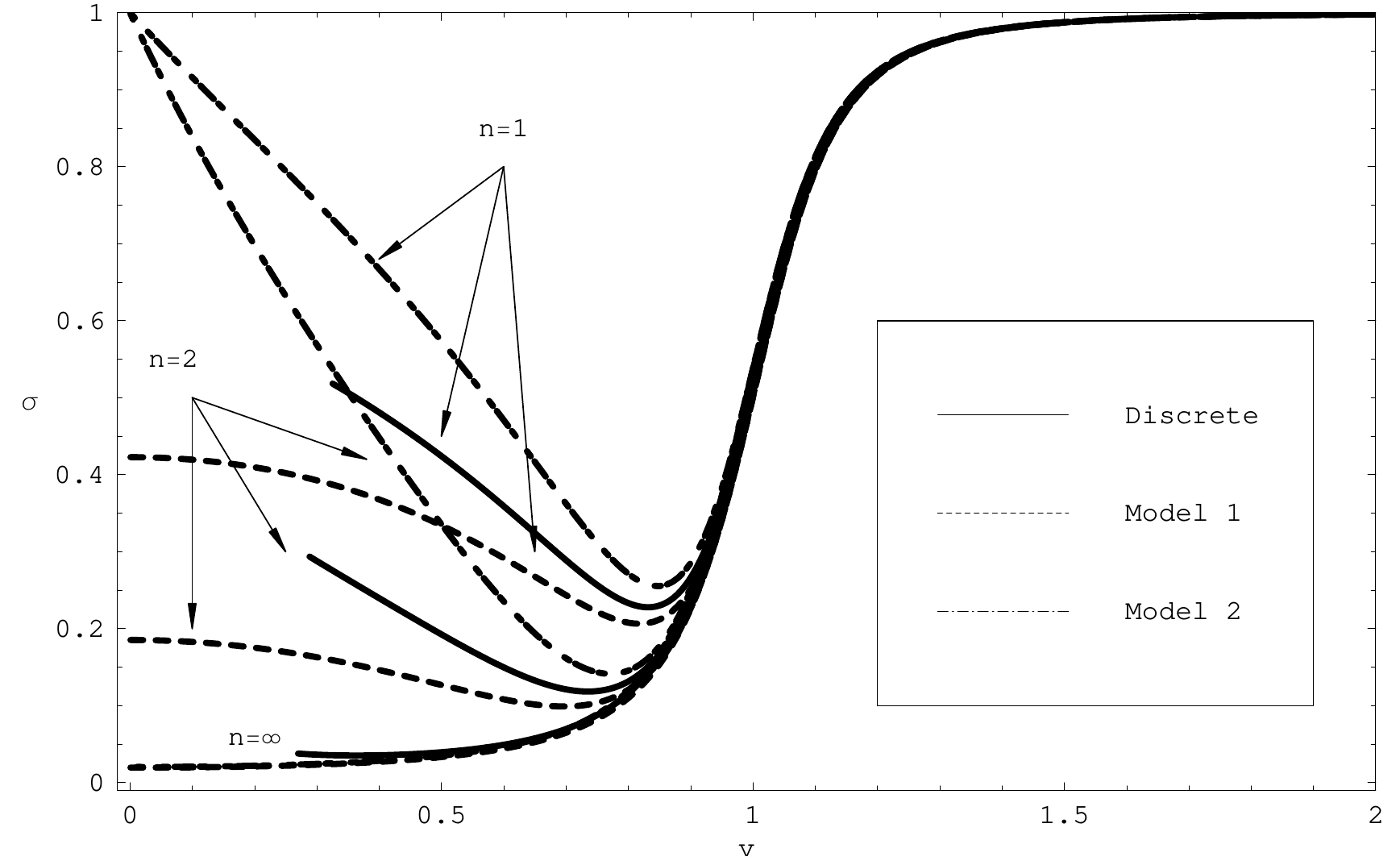}
\end{center}
\caption{Kinetic relations for solitary waves for $\chi=0.5.$}
\label{krall}
\end{figure}

(1) From the solitary wave profiles and $\sigma v$-relations, as shown in Fig. \ref{solwavecompare} and Fig. \ref{krall} respectively, for comparison between lattice model, model $1$ and model $2,$ we can see that these models behave differently in the small velocity regime. For sufficiently large velocities the continuum approximations capture the effects due to discreteness, similar to the case of type $1$ waves \citep{kresse, vedantam}. From a point of view of modelling discrete systems using continuum formulation, these results present an example of travelling waves in a lattice model which are very well approximated by quasicontinuum approximations although may not be captured by all continuum schemes.

(2) Similar to the waves of type $1$ \citep{kresse, disloc_marder}, without proof we conjecture that there exists a lower bound in the velocity, denoted by $v^*_n,$ of such solitary waves. So only those solitary waves with $n$ number of bumps and which satisfy the property $v\geq v_n^*$ propagate in lattice with formal representation (\ref{disc_sol}). An estimate of $v^*_n,$ which we call $\hat{v}_n$, is possible by using a local condition that needs to be satisfied for all such acceptable velocities. This estimate is given by
$$\hat{v}_n:=\inf\{v: u^{\prime }(n \pi/k_0(v))<0, L(k_0(v))=0\}.$$
which gives
\begin{equation}
\hat{v}_n=\inf\{v: i\sum_{\xi \in S_{d}^{-}}(1-e^{-2i\xi (n \pi/k_0(v))})/L^{\prime}(\xi)<0, L(k_0(v))=0\}.
\label{condsecondphase}
\end{equation}
This also yields a lower bound $\hat{v}_\infty$ on the velocity of propagation of travelling waves of type $1$ \citep{disloc_marder, kresse, aig_cham_roth} and we state the results as a conjecture without proof of monotonicity of $\hat{v}_n$ \eqref{condsecondphase} as $n$ increases,
\begin{equation}
\hat{v}_\infty:=\inf_n{\hat{v}_n}=\{v:\sum_{\xi \in S_{d}^{-}}\text{Im } L^{\prime }(\xi)/\lvert L^{\prime}(\xi )\rvert^{2}<0\}.
\label{conddisloc}
\end{equation}
As shown in Fig. \ref{krd}, $n$ increases, $\sigma(\hat{v}_n)$ appears to decrease almost as fast as $\sigma^n_p$. Note that the velocity of solitary waves is unique if $\sigma$ is such that it is greater than $\sigma$ corresponding to $v=\hat{v}_n$ \eqref{condsecondphase}.

At this point we do not know how to complete the picture between $v=0$ and $v=\hat{v}_n$ \eqref{condsecondphase} for each $n.$ This is similar to the question that concerns situation for the Frenkel-Kontorova dislocation at small velocities \citep{disloc_atkinson} and the enigma of kinematic resonances \citep{rogula} in this regime.

\begin{figure}[h]
\begin{center}
\includegraphics[width=\textwidth]{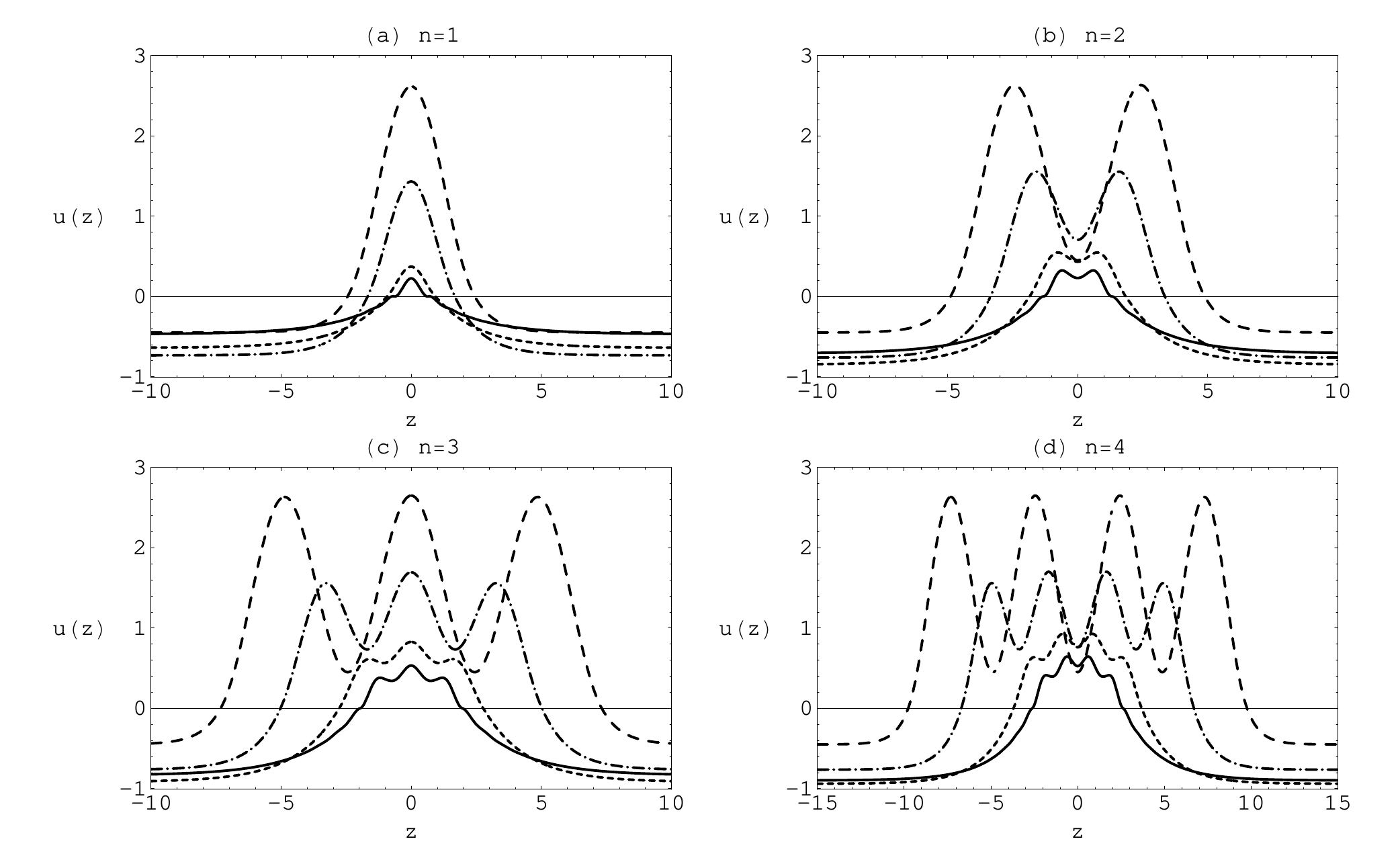}
\end{center}
\caption{Solitary waves in Frenkel-Kontorova Lattice for $\chi=0.5$, same as shown in the pervious figure but arranged with respect to the number of peaks $n$. The solid curves stands for $v=0.3,$ small dashed curve for $v=0.6,$ dash-dot-dash curve for $v=0.9,$ and long dashed curve for $v=1.01.$}
\label{solwaveD2}
\end{figure}

\begin{figure}[h]
\begin{center}
\includegraphics[height=3in]{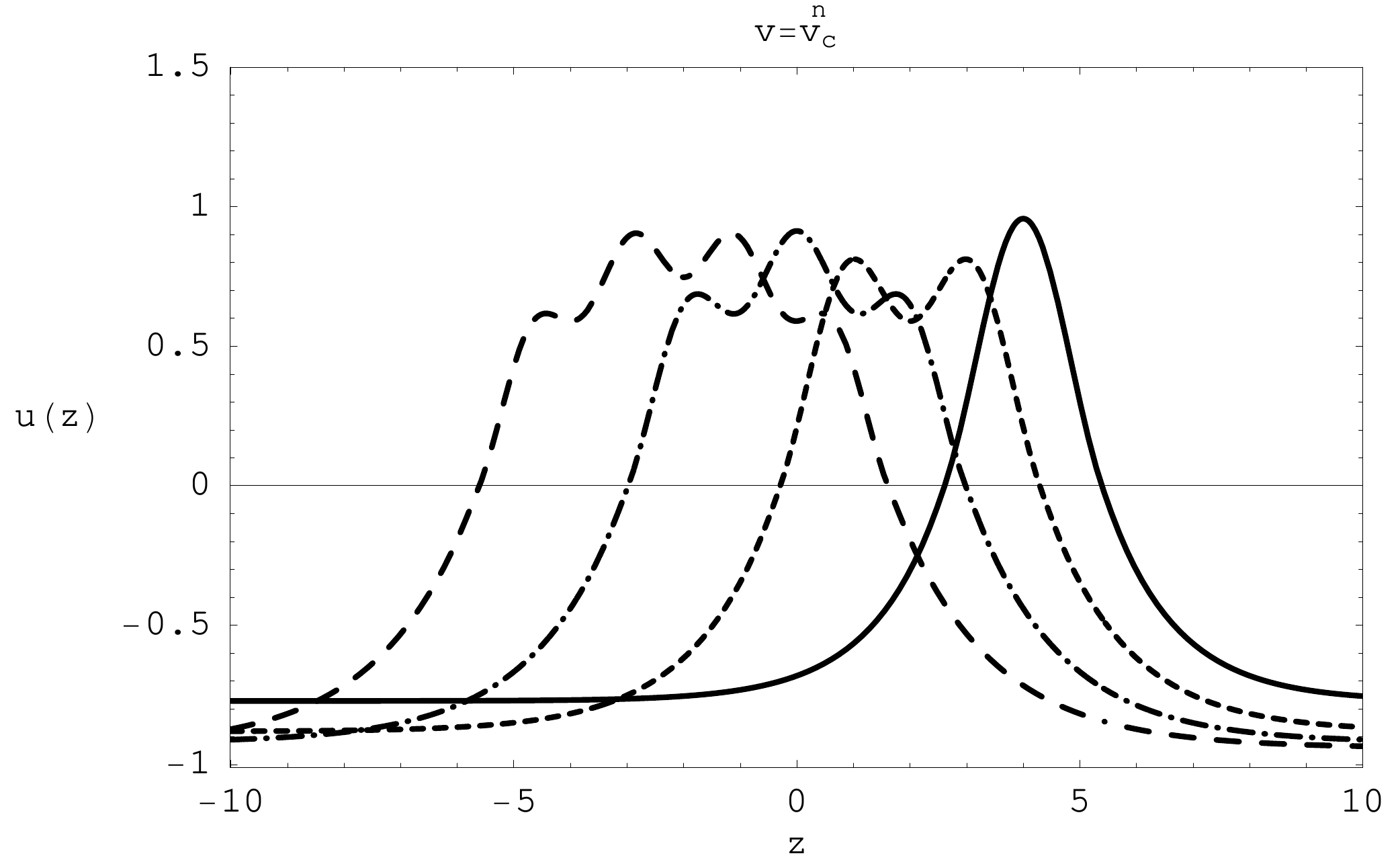}
\end{center}
\caption{Solitary waves at the minimum value of $\sigma$ for $\chi=0.5$ and the case of $n=1, 2, 3$ and $4$ number of bumps.}\label{solwaveD5}
\end{figure}

(3) The solitary wave profiles shown in Fig. \ref{solwaveD} have been also shown in Fig. \ref{solwaveD2}, but they have been grouped with respect to number of bumps in the second phase.
\begin{figure}[h]
\begin{center}
\includegraphics[width=\textwidth]{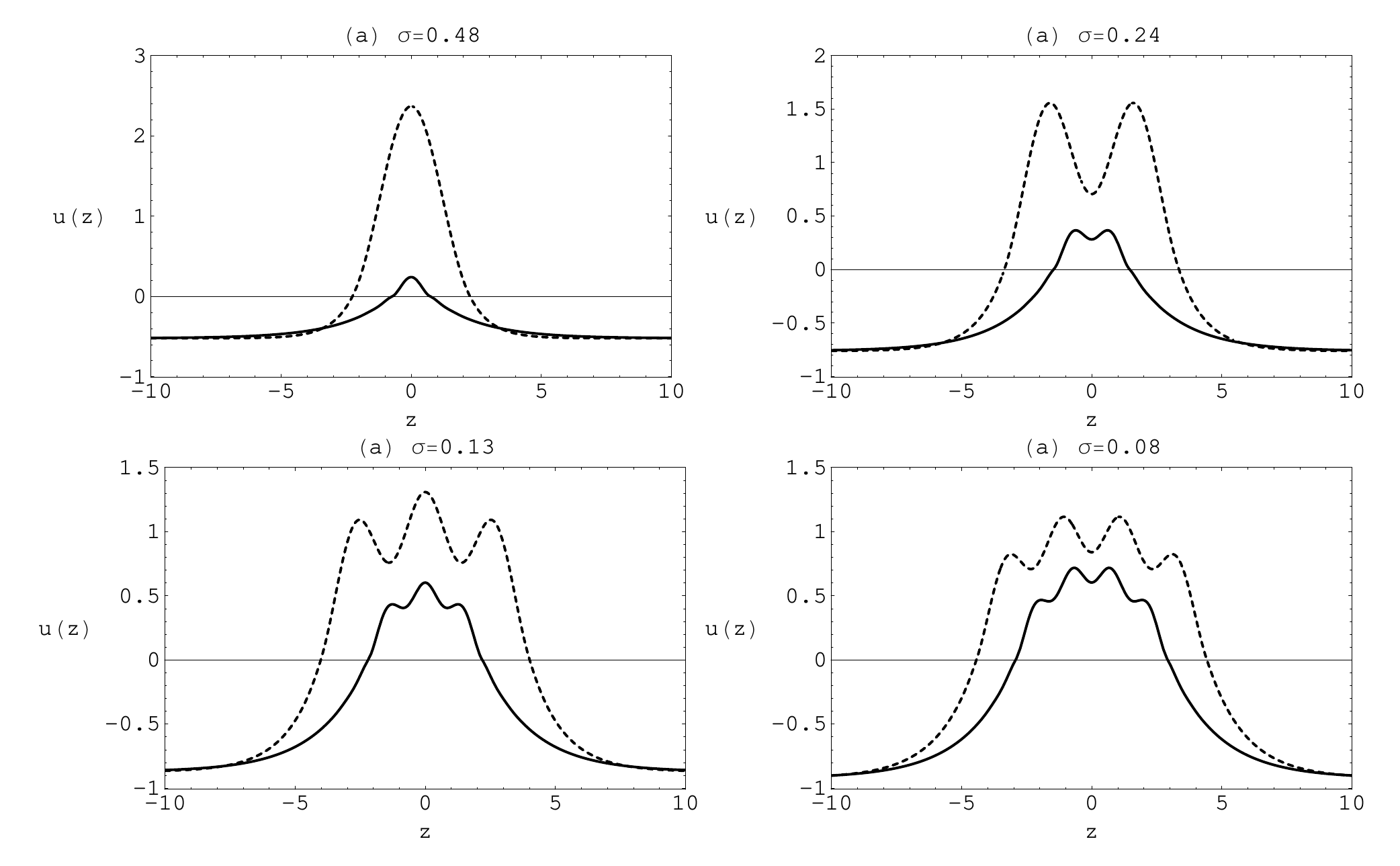}
\end{center}
\caption{Solitary waves for $n=1,2,3,4$ and same $\sigma$ listed, for $\chi=0.5$. Solid curve represents the slower wave and dashed curve represents the faster wave (with a velocity usually around sound speed).}
\label{solwaveD4}
\end{figure}
In Fig. \ref{solwaveD4}, we present the solitary wave profiles possible under the same external stress $\sigma$ and also the same number of bumps. As the number of bumps increases the difference in the two velocities for the same external stress also decreases. For any given $n$ number of bumps, this difference vanishes when $\sigma=\sigma^n_c,$ which is the critical point such that for $\sigma<\sigma^n_c$ the solitary wave with $n$ bumps ceases to exist. Thus $\sigma^n_c$ behaves like the dynamic Peierls stress \citep{weiner7, disloc_atkinson} and reduces to the value for one dislocation in the limiting case of an infinite number of bumps, when effectively the solitary wave reduces to one dislocation or kink. The solitary wave profiles at $\sigma^n_c$ have been shown in Fig. \ref{solwaveD5}, where $v^n_c$ denotes the unique velocity when $\sigma=\sigma^n_c$ for given $n$.

(4) At this point, an observation can been made regarding the relationship between solitary waves and equilibrium states \citep{sharma5}.
\begin{figure}[h]
\begin{center}
\includegraphics[height=3in]{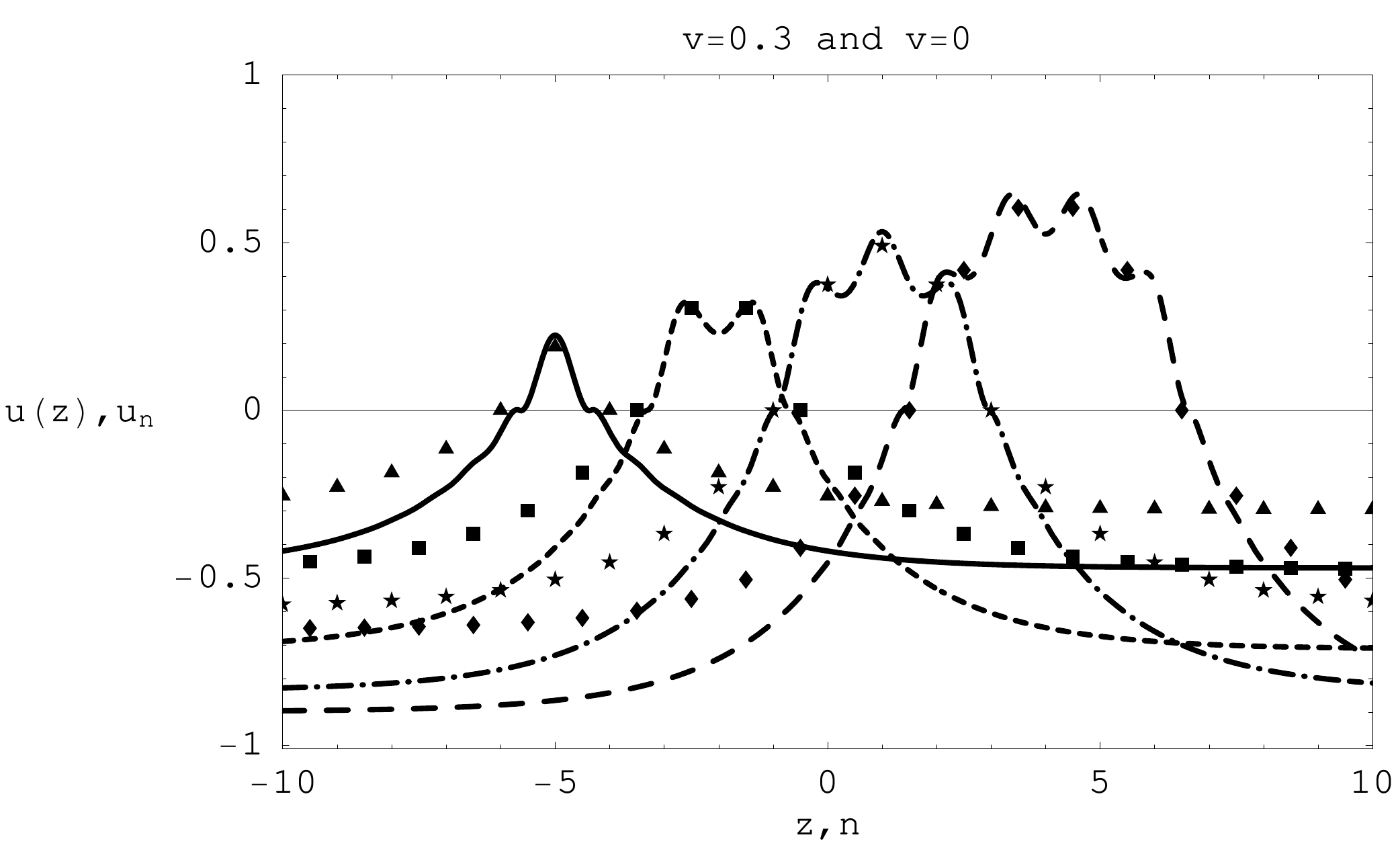}
\end{center}\caption{Solitary waves for $n=1,2,3,4$ for $v=0.3$ and static equilibria \citep{sharma5} for $\chi=0.5$. Solitary wave with $n$ bumps can be identified with an equilibrium state involving $n$ particles in second phase.}\label{solwaveD3}
\end{figure}
We find that $\hat{v}_n$ is close to $0.3$ for $\chi=0.5,$ so using this rough estimate, we have shown the wave profile at $v=0.3$ and compared it with the equilibrium states obtained in the presence of $\sigma=\sigma^n_P$ where $n$ denotes the number of particles in the second phase and $\sigma^n_P$ is the Peierls stress for the corresponding static equilibrium configuration. As shown in Fig. \ref{solwaveD3}, there is one to one correspondence between the number of bumps in second phase of the solitary waves and the number of particles in second phase of the equilibrium states. 

(5) This paper is a result of an accidental observation of solitary waves of type $2$ in continuum approximations of Frenkel-Kontorova lattice. Soon we found that these waves persist in the lattice model and exhibit a rich behaviour in wave profiles as well as $\sigma v$-relations. Recently we were able to gain more insight into the kink-antikink equilibrium states corresponding to the solitary waves of type $2$ \citep{sharma5}. The study of change in free energy \citep{sharma5} associated with transitions between such equilibrium states indicates towards the instability of solitary waves in Frenkel-Kontorova lattice \citep{disloc_hashi}. The instability has been observed when we solved the Euler-Lagrange equations (\ref{discrete model}), numerically, using MATLAB but we just mention these facts here and leave the question for future.

\appendix
\section{Construction of Solitary waves for the Lattice model using Fourier Transforms}
\label{app_dynamics}
Similar to the method adopted in \citep{disloc_atkinson, kresse}, using (\ref{disc_ans1}) we introduce a part of the Fourier Transform, and from (\ref{disc_steady}) we get
\begin{eqnarray}
&&v^{2}\int_{z_{0}}^{\infty }e^{i\xi z}\frac{d^{2}u}{dz^{2}}dz-(\int_{z_{0}}^{\infty }e^{i\xi z}u(z+1)dz-2\int_{z_{0}}^{\infty }e^{i\xi z}u(z)dz \notag\\
&&+\int_{z_{0}}^{\infty }e^{i\xi z}u(z-1)dz)+\chi^{2}[\int_{z_{0}}^{\infty }e^{i\xi z}u(z)dz-(\sigma-1)\int_{z_{0}}^{\infty}e^{i\xi z}dz]=0.\notag
\end{eqnarray}
Using integration by parts and the definitions $L(\xi )=\chi^{2}+4\sin^{2}\frac{\xi }{2}-v^{2}\xi^{2}, \hat{u}_{r}(\xi )=\int_{z_{0}}^{\infty }e^{i\xi z}u(z)dz,$ we get a part of $\hat{u}$,
\begin{eqnarray}
\hat{u}_{r}(\xi ) &=&\frac{1}{\sqrt{2\pi }L(\xi )}[v^{2}e^{i\xi z_{0}}u_{+}^{\prime }(z_{0})-i\xi v^{2}e^{i\xi z_{0}}u_{+}(z_{0})-\int_{z_{0}}^{z_{0}+1}e^{i\xi (y-1)}u(y)dy \notag\\
&&+\int_{z_{0}-1}^{z_{0}}e^{i\xi (y+1)}u(y)dy+\chi^{2}(\sigma-1)\int_{z_{0}}^{\infty }e^{i\xi z}dz].\label{FP1}
\end{eqnarray}
Similarly other parts of Fourier Transform of $u$ with definition, $\hat{u}_{l}(\xi )=\int_{-\infty }^{-z_{0}}e^{i\xi z}u(z)dz,$ $ \hat{u}_{c}(\xi )=\int_{-z_{0}}^{z_{0}}e^{i\xi z}u(z)dz$, are given by
\begin{eqnarray}
\hat{u}_{l}(\xi ) &=&\frac{1}{\sqrt{2\pi }L(\xi )}[-v^{2}e^{-i\xi z_{0}}u_{+}^{\prime }(-z_{0})+i\xi v^{2}e^{-i\xi z_{0}}u_{+}(-z_{0})+\int_{-z_{0}}^{-z_{0}+1}e^{i\xi(y-1)}u(y)dy\notag\\
&&-\int_{-z_{0}-1}^{-z_{0}}e^{i\xi (y+1)}u(y)dy+\chi^{2}(\sigma-1)\int_{-\infty }^{-z_{0}}e^{i\xi z}dz]\label{FP2}
\end{eqnarray}
and
\begin{eqnarray}
\hat{u}_{c}(\xi ) &=&\frac{1}{\sqrt{2\pi }L(\xi )}[-v^{2}e^{i\xi z_{0}}u_{-}^{\prime }(z_{0})+v^{2}e^{-i\xi z_{0}}u_{-}^{\prime }(-z_{0}) \notag\\
&&+i\xi e^{i\xi z_{0}}v^{2}u_{-}(z_{0})-i\xi e^{-i\xi z_{0}}v^{2}u_{-}(-z_{0})-\int_{-z_{0}}^{-z_{0}+1}+\int_{z_{0}}^{z_{0}+1}e^{i\xi (y-1)}u(y)(\xi ) \notag\\
&&+\int_{-z_{0}-1}^{-z_{0}}-\int_{z_{0}-1}^{z_{0}}e^{i\xi(y+1)}u(y)dy+\chi^{2}(2+{\sigma-1})\int_{-z_{0}}^{z_{0}}e^{i\xi z}dz].\label{FP3}
\end{eqnarray}
Combining all three parts (\ref{FP1}), (\ref{FP2}) and (\ref{FP3}), and {\em assuming} continuity and differentiability of $u$ at $z=\pm z_{0}$, we get the full function $\hat{u},$ the Fourier Transform of $u.$ From the inverse Fourier Transform of $\hat{u}$, we obtain
\begin{eqnarray}
u(z) &=&\frac{1}{2\pi }\chi^{2}\int_{-\infty }^{\infty }\frac{e^{-i\xi z}}{L(\xi )}[(\sigma-1)\int_{-\infty }^{\infty }e^{i\xi z}dz+2\int_{-z_{0}}^{z_{0}}e^{i\xi z}dz]d\xi \notag\\
&=&{\sigma-1}+\frac{2\chi^{2}}{\pi }\int_{-\infty }^{\infty }\frac{e^{i\xi (z_{0}-z)}-e^{-i\xi (z_{0}+z)}}{2iL(\xi )\xi }d\xi.
\label{FU}
\end{eqnarray}
Assume $z_{0}\neq 0,$ so there is no pole at $\xi=0.$
Also due to lack of radiative damping as $\left| z\right| \rightarrow \infty,$ we don't expect any elastic waves in the solution for $\left| z\right| >z_{0}.$ Therefore, in order to invoke the residue theorem, we complete the integration along a contour so that all the real valued poles of the integrand are excluded.
\begin{figure}[h]
\begin{center}
\includegraphics[height=2.5in]{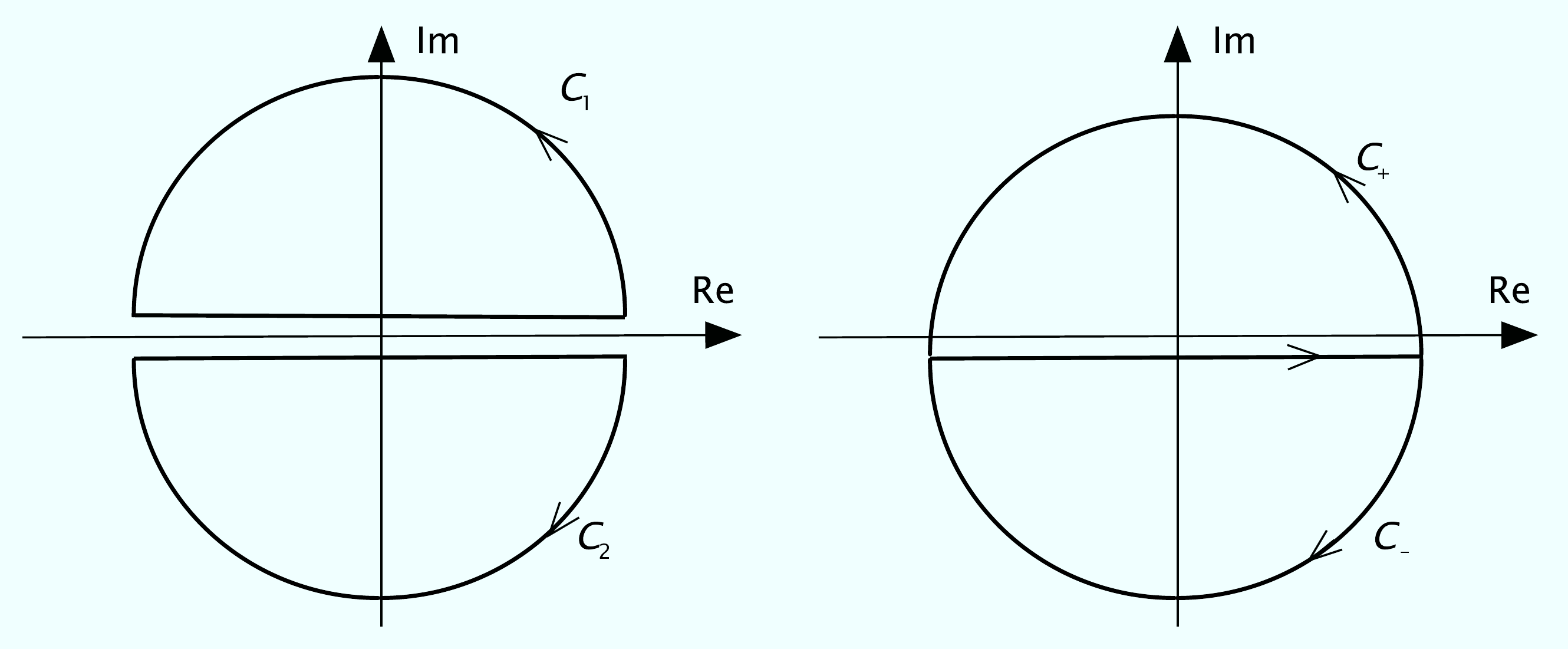}
\end{center}
\caption{Contours $C_1, C_2, C_\pm$ for Integration. The radius is considered to go to infinity in the evaluation of the integrals and the contribution from the peripheral part goes to zero very fast due to the presence of exponentially decaying term in the integrand.}
\label{contour}
\end{figure}

For $z>z_{0},$ we consider a contour that lies in the lower half of the complex plane, $C_2$ as shown in Fig. \ref{contour}, and by using residue theorem, we get from \eqref{FU}
\begin{equation*}
u(z)={\sigma-1}-4\chi^{2}i\sum_{\xi \in S_{d}^{-}}\frac{e^{-i\xi z}\sin(\xi z_{0})}{L^{\prime }(\xi )\xi }, S_{d}^{-}=\{\xi \mid L(\xi )=0;\text{Im }\xi <0\}.
\end{equation*}
For $z<-z_{0},$ we consider a contour that lies in the upper half of the complex plane, $C_1$ as shown in Fig. \ref{contour}, and using the Residue theorem, we get from \eqref{FU}
\begin{equation*}
u(z)={\sigma-1}+4\chi^{2}i\sum_{\xi \in S_{d}^{+}}\frac{e^{-i\xi z}\sin
(\xi z_{0})}{L^{\prime }(\xi )\xi }, S_{d}^{+}=\{\xi \mid L(\xi )=0;\text{Im }\xi >0\}.
\end{equation*}
We require that $u(-z_{0})=0,$ according to the assumption \eqref{disc_ans1}, and it is equivalent to the following claim.

{\bf Claim 1:}
\begin{equation}
\sum_{\xi \in S_{d}^{-}}\frac{e^{-i\xi z_{0}}\sin (\xi z_{0})}{L^{\prime}(\xi )\xi }+\sum_{\xi \in S_{d}^{+}}\frac{e^{i\xi z_{0}}\sin (\xi z_{0})}{L^{\prime }(\xi )\xi }=0. \label{curious_prop1}
\end{equation}

{\bf Proof:}
Since $\xi \in S_{d}^{-}\Leftrightarrow \xi^{\ast }\in S_{d}^{+},$ and $L^{^{\prime }}(\xi )=2(\sin \xi -v^{2}\xi ),$ so $L^{^{\prime }}(\xi^{\ast})=L^{\prime }(\xi )^{\ast },$ we get
\begin{equation*}
\sum_{\xi \in S_{d}^{-}}\{\frac{e^{-i\xi z_{0}}\sin (\xi z_{0})}{L^{\prime}(\xi )\xi }+\frac{e^{i\xi^{\ast }z_{0}}\sin (\xi^{\ast }z_{0})}{L^{\prime}(\xi^{\ast })\xi^{\ast }}\}=\sum_{\xi \in S_{d}^{-}}\{e^{-i\xi z_{0}}\frac{\sin (\xi z_{0})}{L^{\prime }(\xi )\xi }+(e^{-i\xi z_{0}}\frac{\sin(\xi z_{0})}{L^{\prime }(\xi )\xi })^{\ast }\}.
\end{equation*}
Since $\xi \in S_{d}^{-}\Leftrightarrow -\xi^{\ast }\in S_{d}^{-},$ and $L^{^{\prime }}(-\xi^{\ast })=-L^{\prime }(\xi ),$ the claim follows. QED

For $\left| z\right| <z_{0},$ we cannot exclude real valued poles in the integral term of \eqref{FU}. However, we dissect the expression of integral into two integrals
\begin{eqnarray*}
\int_{-\infty }^{\infty }\frac{e^{i\xi (z_{0}-z)}-e^{-i\xi (z_{0}+z)}}{2iL(\xi )\xi }d\xi &=&\int_{-\infty }^{\infty }\frac{e^{i\xi (z_{0}-z)}}{2iL(\xi )\xi }d\xi-\int_{-\infty }^{\infty }\frac{e^{-i\xi (z_{0}+z)}}{2iL(\xi )\xi }d\xi.\\
\end{eqnarray*}
For the first integral, consider the contour $C_+$, as shown in the second of Fig. \ref{contour}, that lies in the upper half of the complex plane but also includes the real axis and for the second integral consider a contour $C_-$, as shown in Fig. \ref{contour}, that lies, strictly, in the lower half of the complex plane. After the breakup of the contour integrals, we do get a contribution from the pole at $\xi=0$ for the first integral. Thus for $\left| z\right| <z_{0},$
\begin{eqnarray*}
u(z)&=&{\sigma-1}+2+2\chi^{2}[\sum_{\xi \in S_{d}^{-}\cup S_{u}}\frac{e^{-i\xi
z}e^{i\xi z_{0}}}{L^{\prime }(\xi )\xi }+\sum_{\xi \in S_{d}^{+}}\frac{e^{-i\xi z}e^{-i\xi z_{0}}}{L^{\prime }(\xi )\xi }], 
\end{eqnarray*}
with $S_{u}=\{\xi \mid L(\xi )=0;\text{Im }\xi=0, \xi\neq0\}.$

We also need that $u(z)\rightarrow 0$ as $z\rightarrow
\pm z_{0}$ with $\left| z\right| \leq z_{0}$. These two conditions can be simplified to an identity proved in the claim below and the condition
\begin{equation}
\sum_{k\in S_{u}}e^{2ikz_{0}}/(L^{\prime }(k)k)=\sum_{k\in S_{u}}1/(L^{\prime }(k)k).
\label{adefz0}
\end{equation}

{\bf Claim 2:}
\begin{equation}
1+\chi^{2}\sum_{L(\xi )=0, \xi\neq0}\frac{1}{L^{\prime }(\xi )\xi }=0,
\label{curious_prop}
\end{equation}
{\bf Proof:}
Consider the contour $C_-$ that lies in the lower half of the complex and the contour $C_+$ that lies in the upper half complex plane including the real axis. Then $\int_{-\infty }^{\infty }\frac{1}{L(\xi )\xi }d\xi=\int_{C_{-}}\frac{1}{L(\xi )\xi }d\xi=-2\pi i\sum_{\xi \in S_{d}^{-}}\frac{1}{L^{\prime }(\xi)\xi },$ but also $\int_{-\infty }^{\infty }\frac{1}{L(\xi )\xi }d\xi=\int_{C_{+}}\frac{1}{L(\xi )\xi }d\xi=2\pi i(\frac{1}{L(0)}+\sum_{k\in S_{u}}\frac{1}{L^{\prime }(k)k}+\sum_{\xi \in S_{d}^{+}}\frac{1}{L^{\prime}(\xi )\xi }).$ Therefore 
\begin{equation*}
\frac{1}{L(0)}+\sum_{k\in S_{u}}\frac{1}{L^{\prime }(k)k}+\sum_{\xi \in
S_{d}^{-}}\frac{1}{L^{\prime }(\xi )\xi }+\sum_{\xi \in S_{d}^{+}}\frac{1}{L^{\prime }(\xi )\xi }=0.
\end{equation*}
Since $L(0)=\chi^{2},$ the identity (\ref{curious_prop}) follows. QED

In the construction of the solitary waves, using the Fourier Transform of $u,$ we assume the continuity of the derivative of $u.$ For the derivative, as $z\searrow z_{0},$
\begin{equation*}
u^{\prime }(z_{0}^{+})=-4\chi^{2}\sum_{\xi \in S_{d}^{-}}\frac{e^{-i\xi z_{0}}\sin (\xi z_{0})}{L^{\prime }(\xi )}=2i\chi^{2}\sum_{\xi \in S_{d}^{-}}\frac{1-e^{-2i\xi z_{0}}}{L^{\prime }(\xi )},
\end{equation*}
and as $z\nearrow z_{0},$
\begin{eqnarray*}
u^{\prime }(z_{0}^{-}) &=&-i2\chi^{2}\sum_{\xi \in S_{u}\cup S_{d}^{-}\cup S_{d}^{+}}\frac{1}{L^{\prime }(\xi )}+i2\chi^{2}\sum_{\xi \in S_{d}^{-}}\frac{1-e^{-2i\xi z_{0}}}{L^{\prime }(\xi )}\\
&=&-\frac{i2\chi^{2}}{2}\sum_{\xi \in S_{u}\cup S_{d}^{-}\cup S_{d}^{+}}\overset{0}{\overbrace{[\frac{1}{L^{\prime }(\xi )}+\frac{1}{L^{\prime }(-\xi )}]}}+i2\chi^{2}\sum_{\xi \in S_{d}^{-}}\frac{1-e^{-2i\xi z_{0}}}{L^{\prime }(\xi )} \\
&=&i2\chi^{2}\sum_{\xi \in S_{d}^{-}}\frac{1-e^{-2i\xi z_{0}}}{L^{\prime }(\xi )}=u^{\prime }(z_{0}^{+}).
\end{eqnarray*}
Thus, the solitary waves are continuously differentiable at $+z_0.$ However we find that the continuity of derivative at $z=-z_0$ imposes a restriction on $z_0$ itself as condition \eqref{condsmooth}. In the following claim we prove this condition as necessary.

{\bf Claim 3:}
Condition \eqref{condsmooth} is necessary for the continuity of derivative of solitary wave.

{\bf Proof:}
As $z\searrow -z_{0},$
\begin{eqnarray*}
u^{\prime }(-z_{0}^{+}) &=&-i2\chi^{2}\sum_{\xi \in S_{u}}\frac{e^{2i\xi z_{0}}}{L^{\prime }(\xi )}-i2\chi^{2}\sum_{\xi \in S_{d}^{+}}\frac{e^{2i\xi z_{0}}}{L^{\prime }(\xi )}-i2\chi^{2}\sum_{\xi \in S_{d}^{-}}\frac{1}{L^{\prime }(\xi )} \\
&=&-i2\chi^{2}\{\sum_{\xi \in S_{u}}\frac{e^{2i\xi z_{0}}}{L^{\prime}(\xi )}+\sum_{\xi \in S_{d}^{+}}\frac{e^{2i\xi z_{0}}}{L^{\prime }(\xi )}+\sum_{\xi \in S_{d}^{-}}\frac{e^{-2i\xi z_{0}}}{L^{\prime }(\xi )}\}-u^{\prime }(z_{0}^{+}).
\end{eqnarray*}
But 
\begin{eqnarray*}
&&\sum_{\xi \in S_{u}}\frac{e^{2i\xi z_{0}}}{L^{\prime }(\xi )}+\sum_{\xi\in S_{d}^{+}}\frac{e^{2i\xi z_{0}}}{L^{\prime }(\xi )}+\sum_{\xi \in S_{d}^{-}}\frac{e^{-2i\xi z_{0}}}{L^{\prime }(\xi )} \\
&=&\sum_{\xi \in S_{u}}\frac{e^{2i\xi z_{0}}}{L^{\prime }(\xi )}+\sum_{\xi\in S_{d}^{+}}(\frac{e^{2i\xi z_{0}}}{L^{\prime }(\xi )}+\frac{e^{-2i\xi^{\ast }z_{0}}}{L^{\prime }(\xi^{\ast })})\text{ }(\xi \in S_{d}^{-}\Leftrightarrow \xi^{\ast }\in S_{d}^{-}) \\
&=&\sum_{\xi \in S_{u}}\frac{e^{2i\xi z_{0}}}{L^{\prime }(\xi )}+\sum_{\xi\in S_{d}^{+}}(\overset{0}{\overbrace{\frac{e^{2i\xi z_{0}}}{L^{\prime }(\xi)}+\frac{e^{2i\xi z_{0}}}{L^{\prime }(-\xi )}}})\text{ }(\xi \in S_{d}^{-}\Leftrightarrow -\xi^{\ast }\in S_{d}^{+}) \\
&=&2i\sum_{\xi \in S_{u}\cap \mathbb{R}^{+}}\frac{\sin (2kz_{0})}{L^{\prime}(k)}.
\end{eqnarray*}
So $z\searrow -z_{0}, u^{\prime }(-z_{0}^{+})=4\chi^{2}\sum_{\xi \in S_{u}\cap \mathbb{R}^{+}}\frac{\sin (2kz_{0})}{L^{\prime }(k)}-u^{\prime }(z_{0}^{+}).$ And as $z\nearrow -z_{0},$
\begin{eqnarray*}
u^{\prime }(-z_{0}^{-}) &=&4\chi^{2}\sum_{\xi \in S_{d}^{+}}\frac{e^{i\xi z_{0}}\sin (\xi z_{0})}{L^{\prime }(\xi )} \\
&=&4\chi^{2}\sum_{\xi \in S_{d}^{-}}\frac{e^{-i\xi z_{0}}\sin (-\xi z_{0})}{L^{\prime }(-\xi )}\text{ }(\xi \in S_{d}^{-}\Leftrightarrow -\xi^{\ast }\in S_{d}^{-}) \\
&=&2i\chi^{2}\sum_{\xi \in S_{d}^{-}}\frac{e^{-2i\xi z_{0}}-1}{L^{\prime }(\xi )}=-u^{\prime }(z_{0}^{+}).
\end{eqnarray*}
Thus the continuity of derivative of $u$ at $-z_0$ implies the restriction \eqref{condsmooth}. QED

\section{Solitary waves in Continuum Models}
\label{app_continuum}
To obtain solitary waves in the continuum model with second order strain energy (\ref{secondsteady}), the method is similar to the one shown here. We now describe the method used to obtain solitary waves for the continuum approximation (\ref{eqnpade22}). Let $u$ be given by the ansatz
\begin{equation}
u(z)={\sigma-1}+\left\{ 
\begin{array}{cc}
\Sigma_{i=1}^{4}A^l_{i}e^{-r_{i}\left| z\right| }, & z<-z_0, \\ 
2+\Sigma_{i=1}^{4}B_{i}e^{-r_{i}z}, & z\in \lbrack -z_{0},z_{0}]\\
\Sigma_{i=1}^{4}A^r_{i}e^{-r_{i}\left| z\right| }, & z>z_0
\end{array}
\right.
\end{equation}
with $r_{1,2}=\pm \sqrt{\frac{-(1-v^{2}+\kappa \chi^2)+R}{2\kappa v^{2}}},$ $r_{3,4}=\pm i\sqrt{\frac{(1-v^{2}+\kappa \chi^2)+R}{2\kappa v^{2}}},$ and\\ $R=\sqrt{(1-v^{2}+\kappa \chi^2)^{2}+4\chi^{2}\kappa v^{2}}.$

With $\lim_{z\rightarrow\infty }u(z)=\lim_{z\rightarrow-\infty }u(z)$ and all derivatives of $u$ approach $0$ as $z\rightarrow\pm\infty,$ and the jump conditions $\llbracket u\rrbracket=\llbracket u_z\rrbracket=\llbracket u_{zzz}\rrbracket=0, v^2\llbracket u_{zz}\rrbracket=2\chi^2$ at $z=\pm z_0,$
\begin{equation}
\begin{array}{ccccccc}
A^r_{1}e^{-r_{1}z_{0}}&=&2+\Sigma_{i=1}^{4}B_{i}e^{-r_{i}z_{0}};& &A^l_{2}e^{-r_{1}z_{0}}&=&2+\Sigma_{i=1}^{4}B_{i}e^{r_{i}z_{0}},\\
-r_{1}A^r_{1}e^{-r_{1}z_{0}}&=&-\Sigma_{i=1}^{4}r_{i}B_{i}e^{-r_{i}z_{0}};& &r_{1}A^l_{2}e^{-r_{1}z_{0}}&=&-\Sigma_{i=1}^{4}r_{i}B_{i}e^{r_{i}z_{0}},\\
r_{1}^2A^r_{1}e^{-r_{1}z_{0}}&=&\Sigma_{i=1}^{4}r_{i}^{2}B_{i}e^{-r_{i}z_{0}}-2\chi^2/v^2;& &-r_{1}^2A^l_{2}e^{-r_{1}z_{0}}&=&\Sigma_{i=1}^{4}r_{i}^{2}B_{i}e^{r_{i}z_{0}}+2\chi^2/v^2,\\
-r_{1}^3A^r_{1}e^{-r_{1}z_{0}}&=&-\Sigma_{i=1}^{4}r_{i}^{3}B_{i}e^{-r_{i}z_{0}};& & r_{1}^3A^l_{2}e^{-r_{1}z_{0}}&=&-\Sigma_{i=1}^{4}r_{i}^{3}B_{i}e^{r_{i}z_{0}}.
\end{array}
\end{equation}
After simplification we get $A^l_{2}=A^r_{1}=A_1, B_1=B_2, B_3=B_4$ and 
\begin{equation*}
A_1=-2\sinh{r_1z_0}\frac{-\chi^2/v^2-r_3^2}{r_3^2-r_1^2}, B_1=e^{-r_1 z_0}\frac{-\chi^2/v^2-r_3^2}{r_3^2-r_1^2}, B_3=\frac{-\chi^2/v^2-r_1^2}{(r_1^2-r_3^2)\cosh r_3z_0},
\end{equation*}
and $ z_{0}=\frac{n\pi }{r_{3}},$ where $n$ is a positive integer.

\printbibliography

\end{document}